\documentclass[11pt,reqno,a4paper,oneside]{article}
\usepackage[dvips,letterpaper]{geometry}
\usepackage[dvips]{graphicx}
\usepackage[utf8]{inputenc}
\usepackage{lscape}
\usepackage{pdflscape}
\usepackage{amsmath,amssymb,amsfonts, mathrsfs, bm}
\usepackage{color}
\usepackage[short,12hr]{datetime}
\usepackage{verbatim}
\usepackage{draftcopy}
\usepackage{listingsutf8}
\usepackage{listings}
\usepackage{framed}
\usepackage{algpseudocode}
\usepackage[Algoritmo]{algorithm}
\usepackage{algorithmicx}
\usepackage[table,xcdraw]{xcolor} 
\usepackage{multirow}
\usepackage{eurosym}
\usepackage{subfigure}
\usepackage{natbib}
\usepackage{lifecon}
\usepackage{authblk}
\usepackage{lipsum}

\definecolor{dkgreen}{rgb}{0,0.6,0}
\definecolor{gray}{rgb}{0.5,0.5,0.5}
\definecolor{mauve}{rgb}{0.58,0,0.82}

\newcommand\blfootnote[1]{%
	\begingroup
	\renewcommand\thefootnote{}\footnote{#1}%
	\addtocounter{footnote}{-1}%
	\endgroup
}

\begin{document}

\title{A public micro pension programme in Brazil: Heterogeneity among states and setting up of benefit age adjustment} 
\author[1,2]{Renata G.~Alcoforado}
\author[1]{Alfredo D.~Egídio~dos~Reis}
\affil[1]{ISEG \& CEMAPRE, Universidade de Lisboa}
\affil[2]{Department of Accounting and Actuarial Science, Universidade Federal de Pernambuco}
\date{}

%
%
%
\maketitle

\blfootnote{Authors gratefully acknowledge the financial support from FCT/MCTES - Fundação para a Ciência e a Tecnologia (Portuguese Foundation for Science and Technology) through national funds and when applicable co-financed financed by FEDER, under the Partnership Agreement PT2020 (Project CEMAPRE - UID/MULTI/00491/2019).}
\blfootnote{\textbf{Special thanks} to the Superintendence of the INSS that provided the data}

\vspace*{-1cm}
\begin{center}
	{\textbf{Abstract}}
\end{center}

Brazil is the 5th largest country in the world, despite of having a “High Human Development” it is the 9th most unequal country. The existing Brazilian micro pension programme is one of the safety nets for poor people. To become eligible for this benefit, each person must have an income that is less than a quarter of the Brazilian minimum monthly wage and be either over 65 or considered disabled. That minimum income corresponds to approximately $2$ dollars per day. This paper analyses quantitatively some aspects of this programme in the Public Pension System of Brazil. We look for the impact of some particular economic variables on the number of people receiving the benefit, and seek if that impact significantly differs among the 27 Brazilian Federal Units. We search for heterogeneity. We perform regression and spatial cluster analysis for detection of geographical grouping. We use a database that includes the entire population that receives the benefit. Afterwards, we calculate the amount that the system spends with the beneficiaries, estimate values \textit{per capita} and the weight of each UF, searching for heterogeneity reflected on the amount spent \textit{per capita}. In this latter calculation we use a more comprehensive database, by individual, that includes all people that started receiving a benefit under the programme in the period from 2nd of January 2018 to 6th of April 2018. We compute the expected discounted benefit and confirm a high heterogeneity among UF's as well as gender. We propose achieving a more equitable system by introducing `age adjusting factors' to change the benefit age.

\vspace*{0.2cm}

\vspace*{0.2cm}




%

\newpage
\section{Introduction and Motivation}
\label{intro}
Brazil is the 5th largest country in the world with 8.5 million $km^2$ and its population in 2017 was 209,288.28 (\citealp{TheWorldBank2018} and \citealp{IBGE2018b}). The Brazilian Gross Domestic Product (GDP) is US\$$1.91\, \times 10^{12}$, making it the 9th highest GDP in the world, according to the \cite{InternationalMonetaryFund2018a}. However, the Brazilian GDP \textit{per capita}  (considering Purchasing Power Parity), is US\$$16.11$ thousand dollars annually, corresponding to the 84th higher position, much lower in the ranking \citep{InternationalMonetaryFund2018}.
The Human Development Index (HDI) is 0.759, see \citep{UNDP2018}, which is considered by the UN a High Human Development (HHD), since it is in the interval 0.7-0.8, resulting in the 79th  higher position in the World, according to the Human Development Report from the United Nations Development Programme \citep{UNDP2018a}.

The aforementioned figures may not reflect the entire country. Indeed, according to the \cite{WorldInequalityDatabase2018}, the 1\% richer has a national income share of 28.3\% and the bottom 50\% share is only 13.9\%. Considering income inequality Brazil is considered the 9th most unequal country in the world \citep{OXFAMBrasil2018}. This leads us to consider that indices and other metrics computed over averages should be considered as inappropriate. In terms of people's income, a special concern should be put on the data distribution tail, particularly on the left tail. 

On the 2014 Edition of The State of Food Insecurity in the World, Brazil celebrated its removal from the United Nations Hunger Map \citep{MinisteriodoDesenvolvimentoSocial2014}. The undernourishment rate in Brazil fell by half from 10.7\% in 2000-02 to less than 5\% in 2004-06. That report revealed that Brazil achieved both the Millennium Development Goal (MDG), target of halving the proportion of its people suffering from hunger, and the stricter World Food Summit (WFS) target of reducing by half the absolute number of hungry people \citep{FAO2014}.

In 2018, Brazil was at risk to go back to the UN Hunger Map (according to the General Director of FAO - Food and Agriculture Organization \citep{DaSilva2018}). So, although on average Brazil is considered to have a High Human Development, the uneven wealth distribution makes that HHD is not a reality for its entire population. 
That leads to the belief that the low-income people need extra protection. Besides, these particular people live in risky, or riskier, environments and are more likely to be unable to cope with a crisis, when compared to the ``average people". 

Our study object focus on a special micro pension programme inserted in the Brazilian national social security system (managed by the INSS, \textit{Instituto Nacional de Segurança Social},  meaning National Social Security Institute). This special programme is called Continuous Provision Benefit (CPB) and is a \textit{care benefit} for low income citizens that did not achieve the necessary criteria for getting the regular (public) pension. \cite{Controladoria-GeraldaUniao2019} attests that this specific programme cost in 2018 less than 1\% of the total INSS expenditure.
	

A citizen to be eligible for this benefit must prove that in his household the monthly income \textit{per capita} is less than a quarter of the Brazilian Minimum Wage, less than 2 Euros per day, approximately, and must be either over 65 years old or disabled. Commonly, only one member of the household receives this benefit, although there may be exceptions. This programme can be considered as microinsurance in the sense of the protection of the low-income people against specific perils and covers a variety of different risks, see \cite{Churchill2006}. In this definition, the word \textit{micro} refers to the target market, instead of referring to \textit{low premiums and low benefits}. 
%
%
This is microinsurance from the government, a public programme, it has a positive impact on the household of the eligible citizen and it truly effects on the economy of some particular areas or cities in the country. For  instance, we underline that these applicant individuals can contribute up to 14 years and 11 months before turning 65 or have contributed for a period of time before becoming disabled, but if they do not reach the minimum requirements they do not receive the regular pension.




This paper is a first study on the matter with this kind of dataset from Brazil, since we work with the entire population of beneficiaries ($4,644,698$ people). Our study takes place in a critical moment in Brazil when some changes are already happening, perhaps significant, regarding social security in general, in which the micro programme is included. Some changes are coming out publicly without any disclosure of a proper/rigorous study. The data we work with here was provided to us in 2018, with the permission to analyse but not to be disclosed. However, in the meantime the conditions regarding the data permission have changed, and we feel there is now some uncertainty about future developments and consultation study of the data.
	
We work indeed with two different databases. The first with the entire population of the programme and the second is a sample of the first with individualized information like sex, date of birth, city of residence, UF (Federal Unit) of residence, start benefit date and others.
With the first database, we look at the programme from one angle, looking for heterogeneity and how some economic factors impact differently in the applications among all 27 Federal Units in Brazil. We are not looking for causality, but a good/proper understanding. The main idea is to show the existence of clear inequalities among the Brazilian Federal Units, which in turn may contribute to solving a serious problem. So, besides looking for the impact of some economic variables, we want to see how they cluster. These techniques are not particularly innovative nor highly sophisticated. However, we found appropriate using fairly simple tools for our purpose.


With the second database, we look at the programme from another angle. We aim to analyse the impact of social protection for the elderly on the Brazilian National Social Security System (INSS). To achieve this goal we separated the beneficiaries by UF and by gender (since women and men have quite different life expectancies). Is this inequality reflected on the ``Expected Discounted Benefit" by beneficiary in the System (INSS)? Is the system fair and, if not, can we make it more balanced?


We focus on calculating the Expected Discounted Benefit of the protection for the elderly. That is, we calculate a discounted value of future benefits, {then we propose a way to improve, or make the system fairer}, for this we use Life Expectancy at 65 years old and also at birth. We do so because on the former we need to compute how much the government needs to estimate the amount reserved for this group, and the latter is due to the inequality and heterogeneity among the Federal Units resulting in quite different proportions of the population reaching the age of 65. To achieve a more homogeneous system, we create two age adjusting factors for the programme that divide by UF and gender and by UF, respectively.

This paper is organized as follows. In Section~\ref{S:lit} we do a short literature review, in Section~\ref{S:data} we present our two databases and do descriptive analysis, in Section~\ref{S:model} we do multiple regression modelling using Box-Cox and Yeo-Johnson models, and do appropriate testing. In Section~\ref{S:cluster} we do cluster analysis. In the next section we compute the Expected Discounted Benefit from our sample in the INSS and show how the system appears to be unfair. In Section \ref{S:af} we propose two Age Adjusting Factors (AAF) in order to achieve a more equitable system. Afterwards, in Section~\ref{S:reform} we exhibit a proposed reform for the Brazilian Social Security {and argue} how it impacts the present and future beneficiaries of this programme. Finally, we write some concluding remarks in Section \ref{S:concl}.
\section{Literature Review}
\label{S:lit}
\begin{figure}[h]
	\center
	\includegraphics[width=7cm]{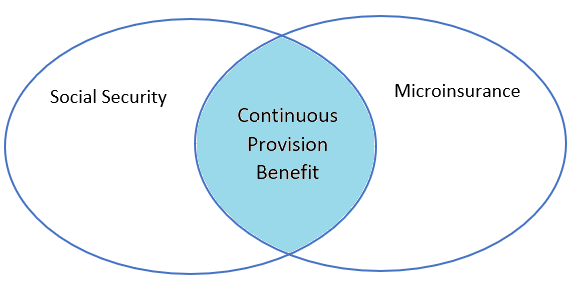}
	\caption{Intersection between Social Security and Microinsurance}
	\label{F:diag}
\end{figure}

In this section we talk about microinsurance in a broader sense, then about the Brazilian social security system - INSS, and finally about the intersection between these two: Microinsurance in the INSS, denoted as ``Continuous Provision Benefit" (directly translated from the Portuguese \textit{Benefício de Provisão Continuada}). Our study is focusing on this intersection. Figure~\ref{F:diag} gives a quick picture of how specific microinsurance interacts with the public pension system in Brazil. 
%
%

As defined in the Introduction, microinsurance is the protection of low-income people (\citealp{Churchill2006}, a reference book on the matter). Poor households often have informal means to manage risks, but informal coping strategies generally provide insufficient protection.
Also, \cite{Jacquier2016} say that
\begin{quotation}
	Microinsurance schemes may assume some social protection functions, such as redistribution through internal cross-subsidies or by channelling public subsidies to their members.
\end{quotation}
	
%
%

From the \cite{INSS2019} we quote (our translation)
\begin{quotation}
	The INSS was created on the 27th of June 1990, through Decree No.~99,350, as a result of the merger of the Institute of Financial Administration of Social Security and Assistance (\textit{Instituto de Administração Financeira da Previdência e Assistência Social}, IAPAS) with the National Institute of Social Security (Instituto Nacional de Previdência Social - INPS), as an {autarchy} linked to the Ministry of Social Security and Social Assistance (\textit{Ministério da Previdência e Assistência Social}, MPAS).
	
	 ...
	 
	The INSS is responsible for the operationalization of the recognition of the rights of the insured individuals of the General Social Security System (RGPS), covering more than 50 million policyholders and having approximately 33 million beneficiaries in 2017. 
	
	...
	
	Article 201 of the Brazilian Federal Constitution observes the organization of RGPS as a contributory and compulsory affiliation, where all the INSS's activities fit in, respecting government policies and strategies from hierarchically superior bodies, such as ministries. The entity is currently linked to the Ministry of the Economy.
\end{quotation} 
%

%
%
\cite{Deblon2012} say that although vulnerability and poverty are not the same, poor people are more vulnerable because they are exposed to a higher number of risks. Therefore, these two reinforce each other. There is a vicious circle: The occurrence of risk decreases people's well-being, it may force them to use their financial and social assets to cope with the effects of such risk, but the vulnerability reduces the ability to extend their economic activities and improve their socio-economic well-being. Social protection is the total set of actions that are carried out by the State or other players to address risk, vulnerability or chronic poverty. In this paper we focus in the State as a player. Social security consider typical risks for people who derive their income from paid labour, i.e., for low-income people. Its goal is to break this vicious circle.

Microinsurance is not a substitute for a social transfer scheme because microinsurance addresses vulnerability rather than chronic poverty, while social transfers provide immediate support to people in poverty. If properly designed, microinsurance constitutes an efficient means of providing individuals in need with social safeguards. In this way, it can potentially contribute to closing existing gaps in coverage with the usual social protection schemes operating in developing countries. There are, however, some limitations to the potential of microinsurance \citep{Deblon2012}.

Apart from Brazil, there are other examples of implementation of programmes of microinsurance for social protection in developing countries, see for instance, \cite{Arun2008}. They outline the status of microinsurance provision in Ghana and Sri Lanka. \cite{Roth2007} attest that the percentage covered by microinsurance in Asia is 2.7 percent, in Africa is 0.3 and in Latin America is 7.8. We restrict our analysis to this sort of microinsurance programme in Brazil, and covers over 4.5 million people. Unlike other authors we do a technical and quantitative  analysis only.

The Continuous Provision Benefit (CPB) of the Organic Law of Social Assistance (LOAS) guarantees a minimum (monthly) wage paid for 12 months (it does not include the traditional 13th salary) for the disabled and for the elderly, 65 years of age or older, who prove not having enough means to provide their own maintenance, or even from his family. It is indeed a \textit{micro pension} programme. Brazilians or Portuguese residing in Brazil can apply to this benefit.

%
\section{The Data and descriptive analysis}
\label{S:data}
Quantitative Information for our study is not easily available, exploring it is pioneering research.  Our data was provided by the Superintendence of the INSS. We realized an existing gap in the literature about this particular subject, we do value the importance of using and analysing this data.
As we said in the Introduction, we have two different sets of data that we are now presenting. The first one consists on the entire population of the programme under study. However this population is only divided by UF and not by gender. On the second database we will present a sample of the population with information on every individual. We will use the first database on Sections \ref{S:model} and \ref{S:cluster} and the second on Sections \ref{S:edb} and \ref{S:af}.

\subsection{The First Database}

The data consists on all the Brazilian citizens (and Portuguese) that receive this specific benefit, the CPB. We also separate the beneficiaries that belong to the group aforementioned, that is, people that receive the benefit called \textit{Amparo Social ao Idoso} and \textit{Amparo Social Pessoa Portadora de Deficiência}, which means ``Social Support to the Elderly" and ``Social Support for Disable Person", respectively, literally translated. The total group is 4,644,698, from these 2,595,775 are disabled and 2,048,923 elderlies,  corresponding to 56\% and 44\% respectively. 
%

Brazil has 27 Federal Units (UF) consisting on 26 states plus the Federal District (the capital district). We refer to Table~\ref{T:codigos} in Appendix A 
where we can see for each UF, names, codes and the different life expectancies that are used throughout this work. 
\begin{figure}[H]
	\centering
	\includegraphics[width=11cm]{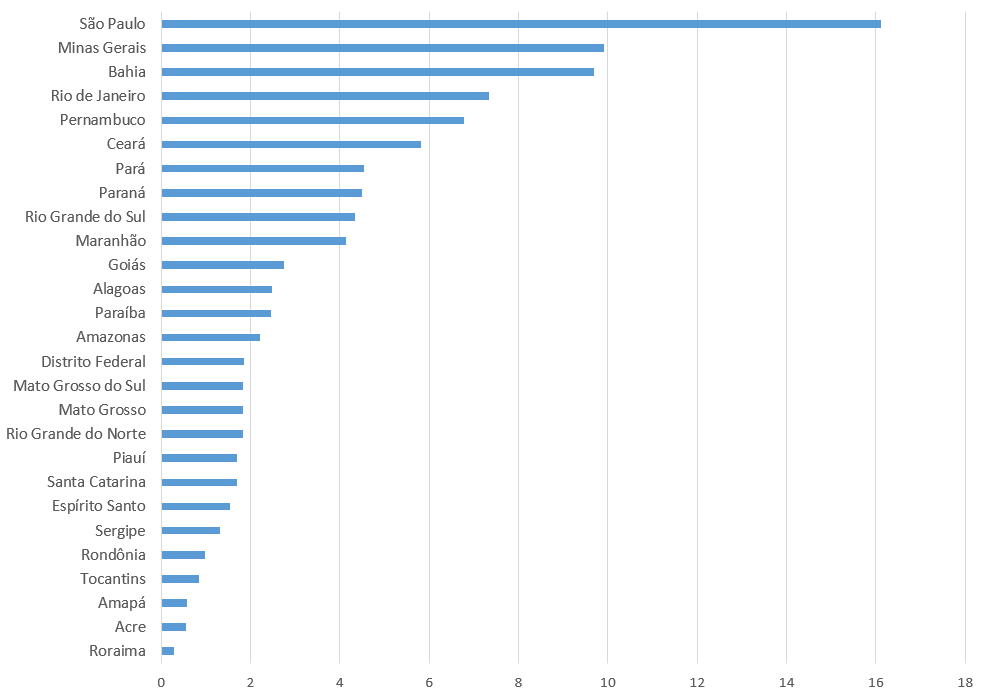}
	\caption{Total bnf (\%)}
	\label{F:col_total}
\end{figure}%

\begin{figure}[H]
	\centering
	\includegraphics[width=12cm]{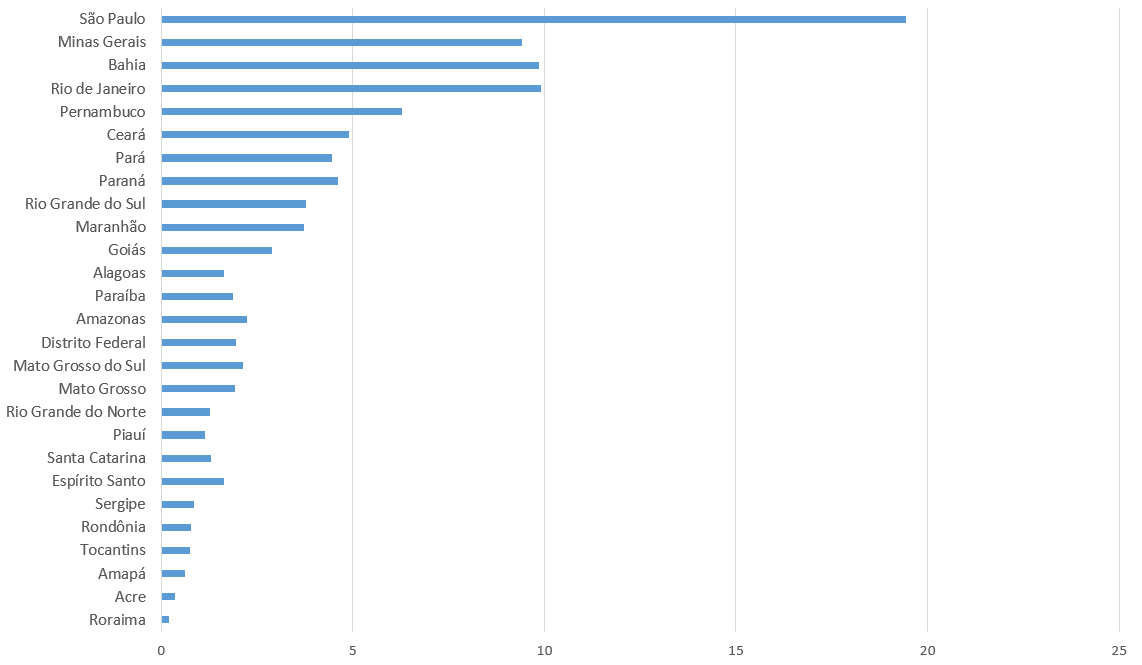}
	\caption{Elderly bnf (\%)}
	\label{F:col_elderly}
\end{figure}

\begin{figure}[H]
	\centering
	\includegraphics[width=12cm]{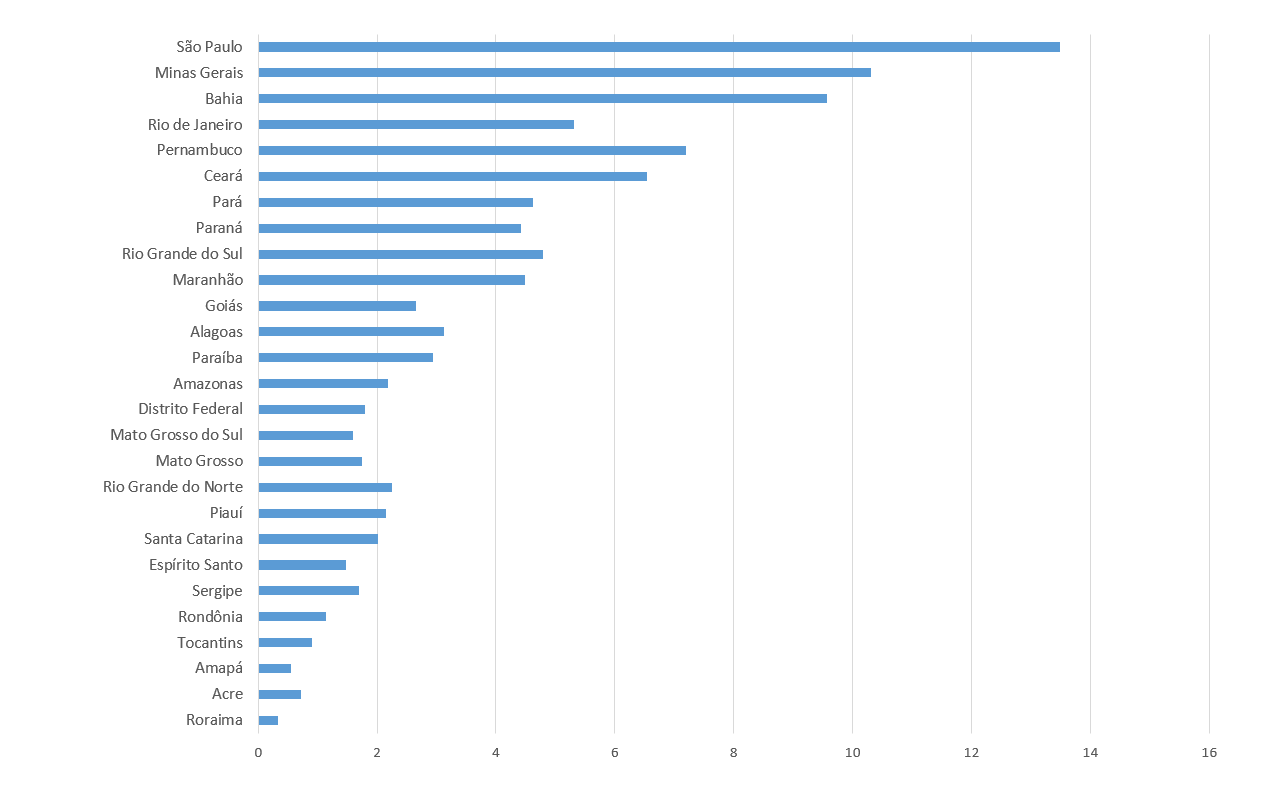}
	\caption{Disabled bnf (\%)}
	\label{F:col_disabled}
\end{figure}%
\begin{figure}[H]
	\centering
	\includegraphics[width=10cm]{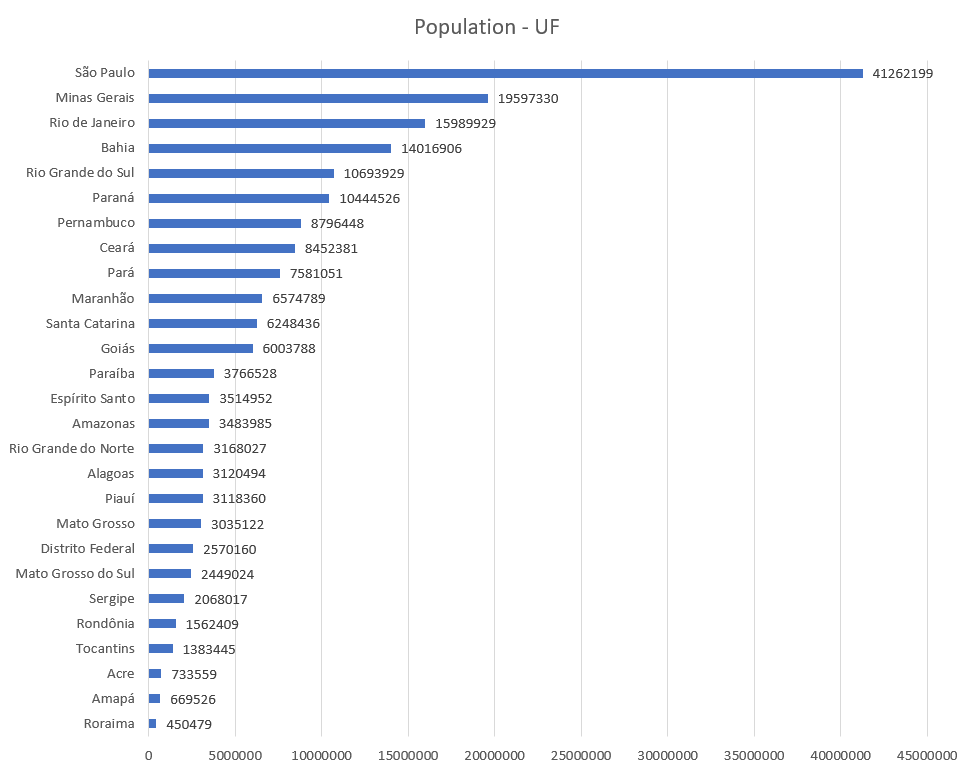}
	\caption{Population by UF}
	\label{F:population}
\end{figure}

The number of beneficiaries vary significantly by UF, but we can not compare this nominal numbers since the UF's can have quite different population sizes. Since each state  population vary from 450,479 to 41,262,199, we found necessary to divide the number of beneficiaries (bnf) by the population size so we can compare ratios from each UF. Figures~\ref{F:col_total}-\ref{F:col_disabled} show the number of beneficiaries by UF, for total, elderly and disabled, respectively. Figure \ref{F:population} refers to the population size.

Figure~\ref{F:col_total} is in descending order by ratio, the following two keep this same order to enhance the position change amongst them. It reveals some interesting points that are worth mentioning, for instance, \textit{Pernambuco} is a state that although is the 5th in the Total ranking, when we consider only Disabled, calls our attention. One of the reasons for this deals with a health problem that occurred in Brazil, \textit{Pernambuco} was particularly affected (microcephaly, 2015).

\begin{figure}[h]
	\center
	\includegraphics[width=13cm]{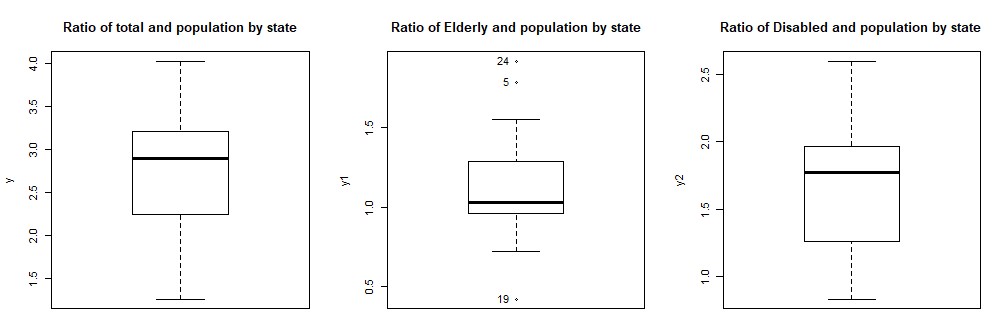}
	\caption{Ratios of the number of people in Social Support and the UF Population}
	\label{F:bp-y}
\end{figure}

In Figure~\ref{F:bp-y} we present three boxplots that show the distribution of the ratios aforementioned. 

When analysing Figure~\ref{F:bp-y}, the first aspect to mention about the shape of the data distribution is that the ratio of the Total and Disabled has a positive asymmetry and the Elderly has a reversed one. Also, the ratio for the Elderly varies from 0.42 to 1.92 while the ratio for the Disabled varies from 0.83 to 2.6, which shows that not only the Elderly is more concentrated but also the scale is lower. 

From these boxplots we can observe that the Elderly group is the one with more prominent outliers. These outliers correspond to Numbers 5, 19 and 24 representing \textit{Mato Grosso do Sul, Santa Catarina} and \textit{Amapá} respectively. Both \textit{Mato Grosso do Sul} and \textit{Amapá} have a lower population density when compared to other states. \textit{Mato Grosso do Sul} is located in the Center-West region, has a density of 6.86 per $km^2$. When we consider the economic variables we analyse, this state does not outstand, indeed it has the 7th highest monthly nominal income \textit{per capita}, R\$~$1,291.00$, and the real average income in (formal) employment is R\$~$2,361.00$, corresponding to the 10th position. 

\textit{Amapá} is located in the North region with a density of $4.69$ per $km^2$, has the second lowest population size among all states, only 669.526 (in 2010 census, \citealp{IBGE2018}) and an estimated population for 2018 of 829.494 people \citep{IBGE2018b}. Although it has the highest ratio for Elderly that are considered poor and receives a benefit, the real average  income  of people in formal employment is R\$~$3,131.00$, which is the second highest average income among the states.

An interesting outlier in this is \textit{Santa Catarina}, that is located in the South region, because this state has the lowest (by far) Elderly receiving benefits and Population ratio, but it has the highest life expectancy (79.1). \textit{Santa Catarina} doesn't have  the highest neither HDI nor income \textit{per capita} but manages to have the lowest ratio. 

\subsection{The Second Database}
\label{S:2data}
The sample that is analysed consisting on citizens that started receiving a pension or a benefit from the INSS in the period from 02/01/2018 until 06/04/2018. We selected the beneficiaries that belong to the group aforementioned, that is, people that receive the benefit called ``Amparo Social ao Idoso" and ``Amparo {Social à Pessoa} Portadora de Deficiência", which means ``Social Support for the Elderly" and ``Social Support for Disable Person", respectively. The total pensioners or beneficiaries of the system consists of 1,332,080. People that received this particular benefit are 81,840, being 40,372 elderlies and 41,468 disabled. 

In Table \ref{T:age} we show a short description of the distribution of the {age at grant of benefit for all beneficiaries in the programme in our sample}: Minimum, 1st to 3rd Quartiles, Maximum and Mean, denoted as $ Min $, $ Q1$-$Q3 $, $ Max $ and $ Mean $, respectively. We removed from the elderly and the disabled group those people who are entitled to the benefit as survivors of previous beneficiaries. We ended up with 40,227 elderlies and 41,387 disabled entries. We note also that often the benefit is granted after the age of $65$ due to a delay in the application process.
\begin{table}[h]
	\begin{center}
	\begin{tabular}{l|cccccc} \hline \hline
		\textbf{Variable} & {$ Min $} & {$ Q1 $} & {$ Median $} & {$ Mean $} & {$ Q3 $} & {$ Max $} \\
		\hline
		\textbf{Disabled} & 0                & 9                    & 33              & 31.45         & 52                   & 81               \\
		\textbf{Elderly}  & 65               & 65                   & 65              & 66.46         & 67                   & 106              \\
		\textbf{Total}    & 0                & 33                   & 64              & 48.71         & 65                   & 106     \\ \hline \hline        
	\end{tabular}
\end{center}
	\vspace{2mm}
	\caption{Distribution of age at grant of benefit (data period: 02/01-06/04/2018)}
	\label{T:age}
\end{table}

\begin{figure}
	\centering
	\includegraphics[width= \textwidth]{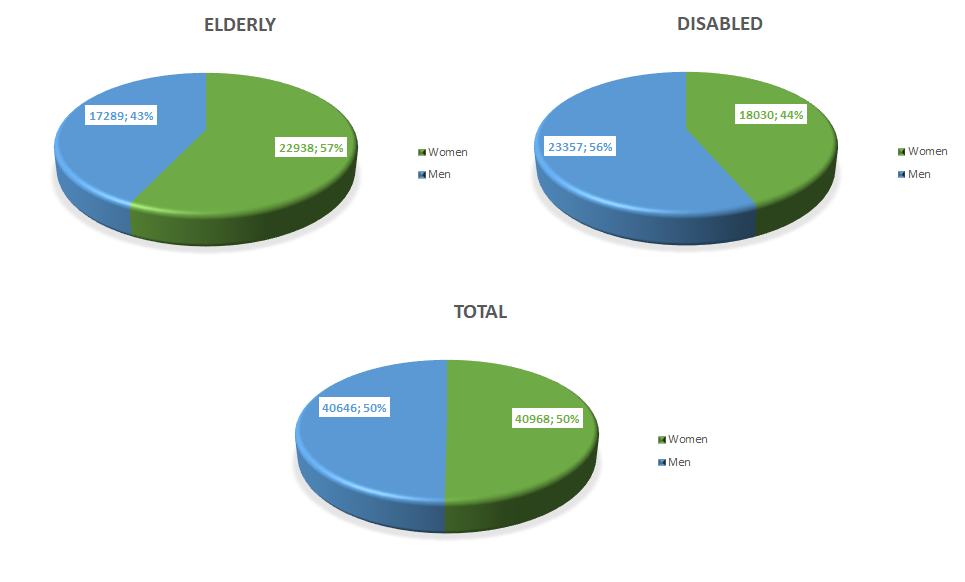}
	\caption{{Distribution of  beneficiaries by sex}}
	\label{F:sex}
\end{figure}

In Figure \ref{F:sex} we present graphs of the distribution of beneficiaries by sex. The first plot corresponds to the Elderly, from those 57\% are women. In the case of the disabled we have 56\% male and among total beneficiaries of CPB of our sample we have 50\% each, approximately. This is probably due to a higher male mortality rate and disability rate. Although in total the distribution looks well balanced, it is not the case when we separate Elderlies and Disabled, where there is a clear gender difference. 

We have available official figures for Life Expectancies by Federal Unit, as well as by gender, for the entire population and we will try to adjust the benefits according to age. These Life Expectancies are displayed in Appendix A. Since the disabled group does not have a minimum age and there is no official Life Expectancy for the disabled population subset, in what follows we devote our efforts on the Elderly Group only. 
\section{Modelling Social Support using Multiple Regression}
\label{S:model}
In this section we aim to study how economic factors can explain the quantity of people receiving the social support of the microinsurance programme in Brazil. 
\cite{Radermacher2012} state that impact evaluations for microinsurance are often complicated, expensive and sometimes difficult to implement. Therefore, less robust techniques predominate. Among the robust approaches  are those that use statistical or econometric techniques.  They also  suggest in particular a regression analysis to take into account of mitigating factors such as income, race or gender. Following these ideas we use income as mitigating factor but not separating people regarding race or gender.
Our quantitative study use regression models where explanatory variables explain the effect by UF/state. 

We chose three different response variables to check whether we find different impacts on them. These three variables are: Total of Social Support, Social Support for the Elderly and Social Support for the Disabled. For each response variable we have 27 entries (UF's). 
The explanatory variables are: HDI (according to the \cite{PNUD2018}), Nominal Monthly Income \textit{per capita} (according to the Ministry of Labour, RAIS), Life Expectancy at birth and Demographic Density (\cite{IBGE2018} and \cite{IBGE2018a}). Since the requirements to acquire the benefits do not distinguish gender, on the regression model we cannot select groups accordingly.

In order to analyse the number of beneficiaries by UF, we consider the above four economic variables, and with these we have set four studying hypotheses, labeled H1-H4:
\begin{itemize}
	\item[H1] We are studying a microinsurance programme, so we start by considering that to an UF with a higher Human Development Index should correspond to a lower ratio on people in need of this programme;
	\item[H2] The main condition to acquire this benefit is to have an income of less than a quarter of the Brazilian minimum monthly wage. So, one of the variables that we use is the Nominal Monthly Income \textit{per capita}, under the hypothesis that UF's with a higher nominal monthly income \textit{per capita} should correspond to a lower ratio of beneficiaries;
	\item[H3] Since a person that acquires this benefit will receive it until death, our third hypothesis is that Federal Units with higher Life Expectancy should result in a higher ratio of beneficiaries. We are going to check if this is replicated on our population;
	\item[H4] Areas that have a higher Demographic Density also present more poverty [see \citealp{Szwarcwald1999} next], so our last hypothesis is that the Federal Units with higher demographic density should also correspond to a higher ratio.
\end{itemize}
\cite{Szwarcwald1999} show that the income inequality affects the homicide rates and this was found precisely in the part of the city with the highest concentration of slum residents (demographic density). In Brazil the following three factors go hand to hand: 1. Income inequality; 2. Demographic density; and 3. More violence which leads to a decrease in Life Expectancy.

\cite{Bourguignon2003} attests that poverty is a multidimensional concept and that to a person to be considered poor it is necessary to fall below at least one of various lines. When applying this concept to Brazil, they define poverty according to income and education. The authors also say that this concept goes in the same direction as the Human Development Index that aggregates \textit{per capita} real GDP, Life Expectancy and Educational Attainment Rate. For this reason, we consider that once more the income monthly \textit{per capita} and Human Development Index are important variables to be taken.

Two years later, \cite{Thorbecke2005} said that authors when trying to measure multi-dimensional poverty only deal with a maximum of four factors and usually use only two. In this paper we used four different factors. \cite{Thorbecke2005} also said that the standard way of assessing whether a person is under or above the poverty line is his income and this is a limited perspective. In this paper we try to see if the four aforementioned factors are also statistically significant for the ratio of beneficiaries of this micro programme.

More recently, \cite{Golgher2016} also talks about multidimensional poverty in Brazil. The author talks about deprivation in households and how some types of deprivation, as food, specially affected the low-income ones. And then how to differentiate medium-income from higher income households from the deprivation of education and some non-popular goods, respectively. In this paper we focus on the low-income households, on those people that are deprived of food, the most basic and important need.

To select a model, we use the four different approaches: 1. Linear; 2. Quadratic (allowing the explanatory variables to be quadratic); 3. Box-Cox transform; and 4. Yeo-Johnson transform (both  are  transformations on the response variable). We take as selection criteria AIC  and BIC (Aikaike and Bayesian Information Criteria, respectively). If both give a similar result we keep the simplest.
At first, the models were selected so that all the included variables and the model itself were statistically significant, so we started with 12 models. Then, we set a first filter that is the Ramsey Regression Equation Specification Error Test - Reset Test \citep{Ramsey1969}. We were left with eight models. Throughout the paper we used a significance level  of 10\%.
Table~\ref{T:AIC} presents the AIC and BIC for the models, those which failed the Reset Test are represented with an ``X" and the chosen models are written in bold. Both criteria lead us to the same conclusion: The model for the Total of beneficiaries is the one with the Yeo-Johnson transform for the response variable. For the Elderly, it is the one with the Box-Cox transform and for the Disabled is the one with no transform on the response variable. All of them  are allowing the explanatory variables to be quadratic.
\begin{table}
	\begin{tabular}{lccc|ccc} \hline \hline
		\textbf{}            & \multicolumn{3}{c}{\textbf{AIC}}                        & \multicolumn{3}{c}{\textbf{BIC}}                           \\ \hline
		\textbf{}            & \textbf{Total}   & \textbf{Elderly} & \textbf{Disabled} & \textbf{Total}     & \textbf{Elderly}  & \textbf{Disabled} \\ 
		\textbf{Linear}      & X                & X                & -54.2             & X                  & X                 & 29.60384          \\
		\textbf{Quadratic}   & -46.63           & -66.68           & \textbf{-64.63}   & 39.77159           & 18.4178           & \textbf{21.76626} \\
		\textbf{Box-Cox}     & -108.96          & \textbf{-73.71}  & X                 & -22.56316          & \textbf{11.39623} & X                 \\
		\textbf{Yeo-Johnson} & \textbf{-133.04} & -72.92           & X                 & \textbf{-46.63863} & 12.1794           & X            \\    \hline \hline
	\end{tabular}
\vspace{1cm}
	\caption{Model selection}
	\label{T:AIC}
\end{table}

The estimated model for the \textbf{Total} group is shown below:
%
\begin{equation*}
Y_{yj} = -8.623 \times 10 - 8.41 \times 10^{-4} X_2 + 2.372 X_3 + 2.942 \times 10^{-7} X_2^2 - 1.605 \times 10^{-2} X_3^2 \, ,
\end{equation*}
where $Y_{yj}$ is the Yeo-Johnson transform of $Y$ (ratio of people by state that receive Social Support),
$X_2$ is the Nominal Monthly Income \textit{per capita} and $X_3$ is the Life Expectancy at Birth.
This model shows a Determination Coefficient - $R^2$ of $0.8007$ that means that it explains about 80\% of the reality that represents, the highest from the three select models (the Adjusted Determination Coefficient, $R^2_a$, is 0.7644). 

Then, we  performed the Bera-Jarque test to check normality in the residuals \citep{Jarque1980}, the {$p$}-value was $0.9797$, resulting in not rejecting a Normal distribution. 
To test linearity we used the Rainbow test \citep{Utts1982}, in this case the {$p$-}value is $0.1802$. So, we do not reject  linearity. To detect heteroskedasticity, we used the Koenker test \citep{Koenker1981}, and we did not reject homoscedasticity, we got  a {$p$-}value of $0.442$.
With respect to multicollinearity, for all three cases using Variance Inflation Factor, we observed multicollinearity, but when checking the correlation between the explanatory variables, the only ones that were high were those  between $X_2$ with $X_2^2$ and $X_3$ with $X_3^2$, which was already expected and we did not considered it a problem.

In Figure~\ref{F:mod} in the Appendix, we present four plots, the first is residuals \textit{versus} fitted values where we do not see any pattern in the residuals. The second one is the Normal Q-Q plot where we can see three outliers, UF's 5, 7 and 19. The third one shows the Scale-Location and the fourth Residuals \textit{versus} Leverage.
In this model we have four outliers, 5, 7, 8, 19, 22 that represent the following federative units: \textit{Mato Grosso do Sul, Goiás, Maranhão, Santa Catarina} and \textit{Distrito Federal}, respectively. From those, \textit{Santa Catarina} and \textit{Distrito Federal} are influential points and 8, 19 and 22 are high-leverage points (using influence measures).

\begin{figure}
	\center
	\includegraphics[width=13cm]{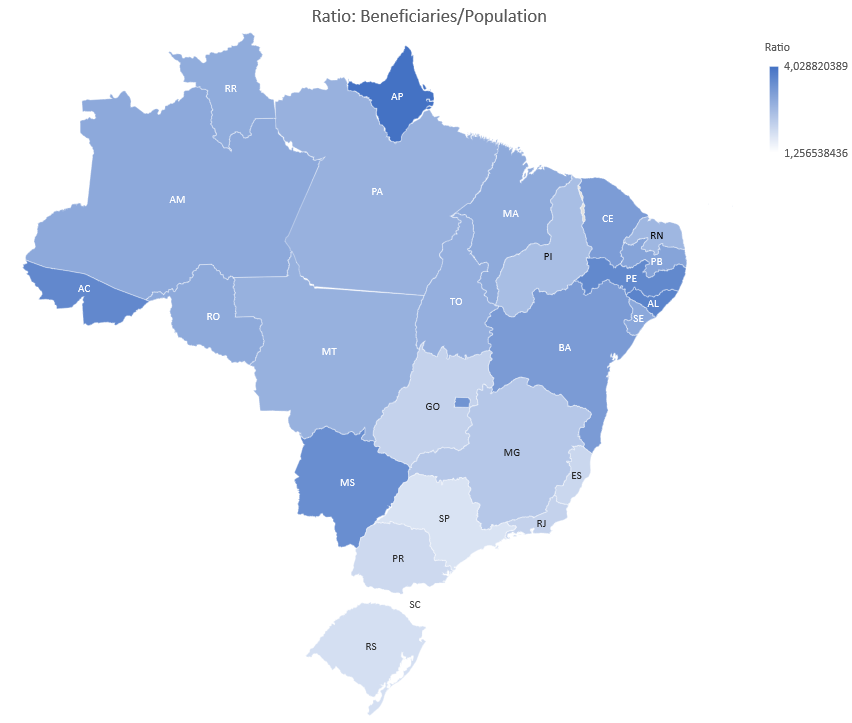}
	\caption{Ratio: Total bnf / Population by UF}
	\label{F:map_total}
\end{figure}

Figure~\ref{F:map_total} show Brazilian map to highlight the UF's that are outliers. The scale represent the range of the ratio, the darker color represent higher values.

\begin{figure}[H]
	\centering
	\begin{minipage}{.5\textwidth}
		\centering
		\includegraphics[width=.9\linewidth, height=150px]{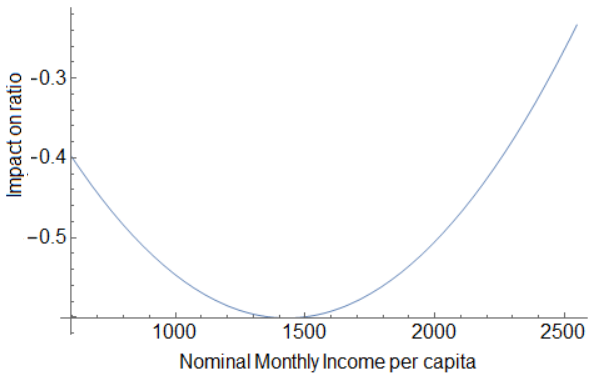}
		\caption{Nominal monthly income on $Y_{YJ}$}
		\label{F:x2-ytotal}
	\end{minipage}%
	\begin{minipage}{.5\textwidth}
		\centering
		\includegraphics[width=.9\linewidth, height=150px]{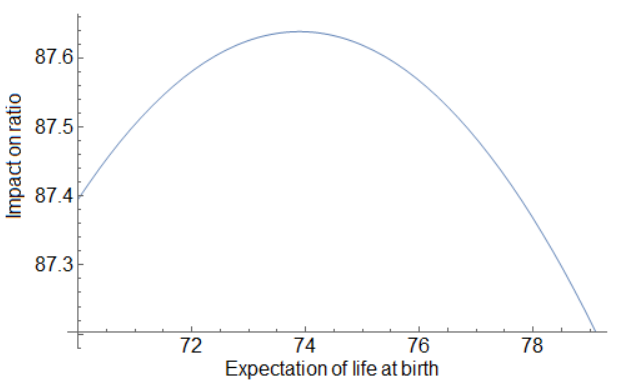}
		\caption{Expectation of life on $Y_{YJ}$}
		\label{F:x3-ytotal}
	\end{minipage}
\end{figure}

Figure~\ref{F:x2-ytotal} and \ref{F:x3-ytotal} present the impact from $X_2$ and $X_3$ on the response variable, the ratio between total bnf and population size.
%
%
We can see that both variables have a linear and a quadratic component. It results on a curved line. In Figure~\ref{F:x2-ytotal} we can see a \textit{smile} shape while in Figure~\ref{F:x3-ytotal} we see the \textit{sad} shape.

The nominal monthly income \textit{per capita} varies from R\$~$597$ to R\$~$2,548$. Until turning point R\$~$1,429$ it decreases the ratio of total beneficiaries by UF. After that it starts to increase, however throughout  the range of income \textit{pc} has always a negative impact.
Life Expectancy at birth has the opposite behaviour. Increases the ratio until $73.9$ years old and after this turning point it starts to decrease. Also, inside the range  the impact is always positive.
%

Focusing now on the Elderly group,  $Y1$ is  now the response variable to the ratio of people by state that receives the Social Support for the Elderly we have to following estimated model:
\begin{equation*}
Y1_{bc} = -1.780 \times 10^2 + 4.837 X_3 + 1.403 \times 10^{-7} X_2^2 - 3.284 \times 10^{-2} X_3^2\, ,
\end{equation*}
where $Y1_{bc}$ is the Box-Cox transform of $Y1$, $X_2$ is the Nominal Monthly Income \textit{per capita} and $X_3$ is the Life Expectancy at birth.

This second model gives an $R^2$ of $0.4557$,  the lowest from all three selected models, it explains almost half of the analysed data. The Adjusted Determination Coefficient, $R^2_a$,  has a value of $0.3847$. 
The Bera-Jarque test has {$p$-}value of 0.6052, and the Normality  was not rejected. The Rainbow test has a {$p$-}value equal to 0.259, so, we do not reject linearity. With respect to the Koenker test we also did not reject the null hypothesis ($p$-value of 0.727).

Figure~\ref{F:mod1} present similar plots to those in Figure~\ref{F:mod} regarding the second model.
As we can see from Figure \ref{F:mod1}, this model presents six outliers, 5, 6, 19, 21, 22 and 24 that represent the following UF's: \textit{Mato Grosso do Sul, Espírito Santo, Santa Catarina, Sergipe, Distrito Federal} and \textit{Amapá}, respectively. The high influential points are \textit{Espírito Santo, Santa Catarina} and \textit{Distrito Federal} and \textit{Distrito Federal} is a high-leverage point.

\begin{figure}[h]
	\center
	\includegraphics[width=13cm]{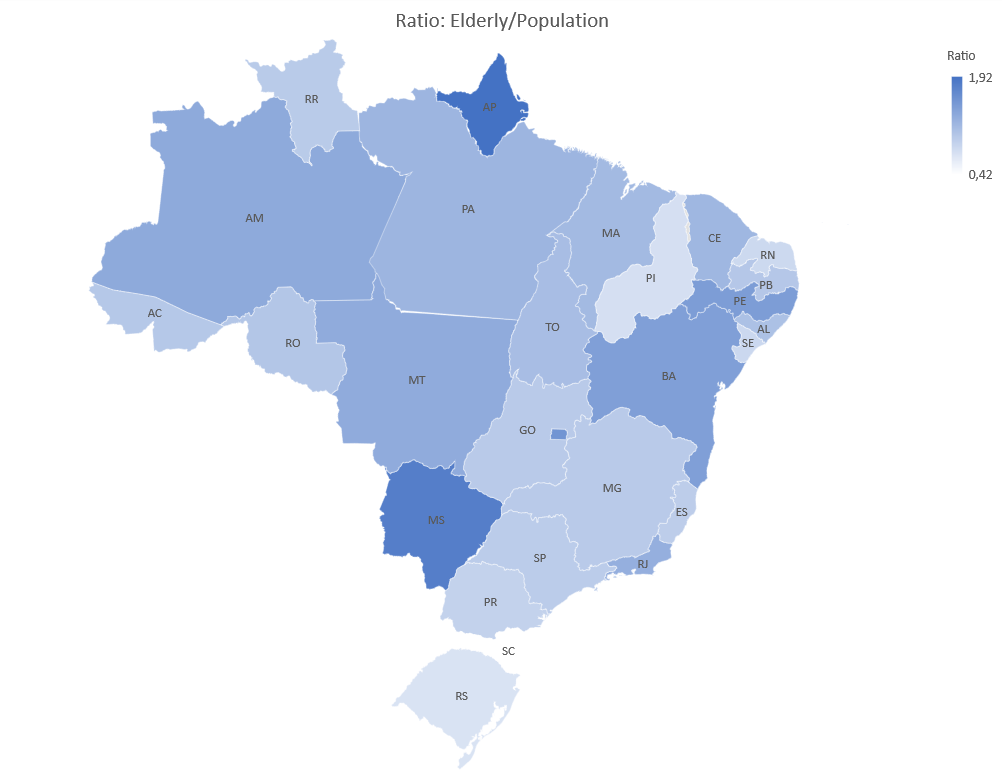}
	\caption{Ratio: Elderly bnf / Population by UF}
	\label{F:map_elderly}
\end{figure}

\begin{figure}[H]
	\centering
	\begin{minipage}{.5\textwidth}
		\centering
		\includegraphics[width=.9\linewidth, height=150px]{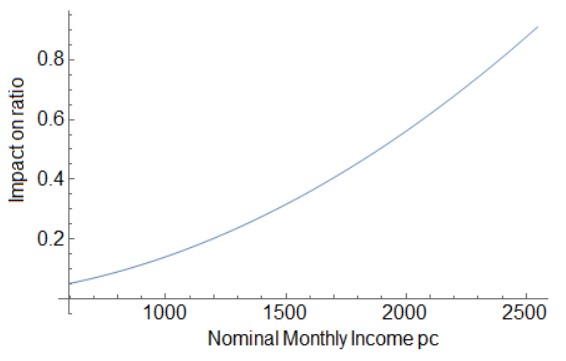}
		\caption{Nominal monthly income on $Y1_{BC}$}
		\label{F:x2-yelderly}
	\end{minipage}%
	\begin{minipage}{.5\textwidth}
		\centering
		\includegraphics[width=.9\linewidth, height=150px]{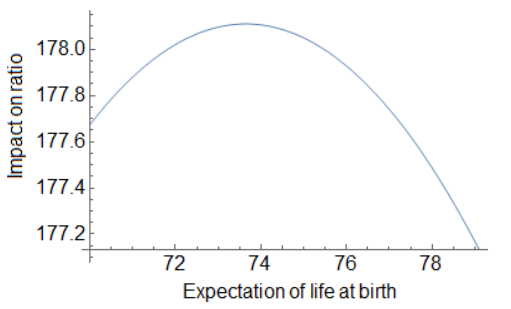}
		\caption{Expectation of life on $Y1_{BC}$}
		\label{F:x3-yelderly}
	\end{minipage}
\end{figure}

In Figure~\ref{F:map_elderly} we show the map with $Y1$, the ratio for the Elderly.
Figures~\ref{F:x2-yelderly} and \ref{F:x3-yelderly} present the impact of $X2$ and $X3$ in $Y1_{bc}$.
%
%
Regarding this model, the variable $X_2$ has only a quadratic component, and $X_3$ has both linear and quadratic. For the entire range, the Nominal Monthly Income \textit{per capita} increases the ratio of the Elderly bnf by UF. Also, this component is always positive.
The Life Expectancy at birth has a different behaviour. It takes the ratio to increase until 73.6 years old and after this turning point it starts decreasing. Also, inside the range of Life Expectancy, the impact is always positive.  

%
 $Y2$ represents the response variable to the number of people by UF that receive the Social Support for Disability we have to following estimated model:
\begin{equation*}
Y2 = -1.445 \times 10^2 - 3.368 \times 10^{-3} X_2 + 4.003 X_3 + 1.018 \times 10^{-6} X_2^2 - 2.696 \times 10^{-2} X_3^2 \, ,
\end{equation*}
where $Y2$ is the Ratio for Disabled, $X_2$ is the Nominal Monthly Income \textit{per capita} and $X_3$ is the Life Expectancy at birth.

At last, on this model the $R^2$ is equal to 0.7207. The $R^2_a$ has a value of $0.6699$.
In this case, similar to the others, both tests did not reject the null hypothesis ($p$-values were $0.7397$ and $0.2524$) for the Bera-Jarque test and the Rainbow test, respectively. The same happened with the test to determine if our variance was heteroskedastic. With a $p$-value of $0.257$, we do not reject the homoscedasticity hypothesis.

Figure~\ref{F:mod2} presents the four plots regarding the model as aforementioned.
On this model, we can observe that 1, 6, 8, 11, 16, 19, 22, 23 and 26 are outliers, they represent \textit{Alagoas, Maranhão, Pará, Rio de Janeiro, Distrito Federal, Acre} and \textit{Roraima}, respectively. The high influential points are \textit{Maranhão, Distrito Federal }and \textit{Roraima} and the high leverage points are \textit{Espírito Santo, Santa Catarina} and \textit{Distrito Federal}. 

\begin{figure}[h]
	\center
	\includegraphics[width=13cm]{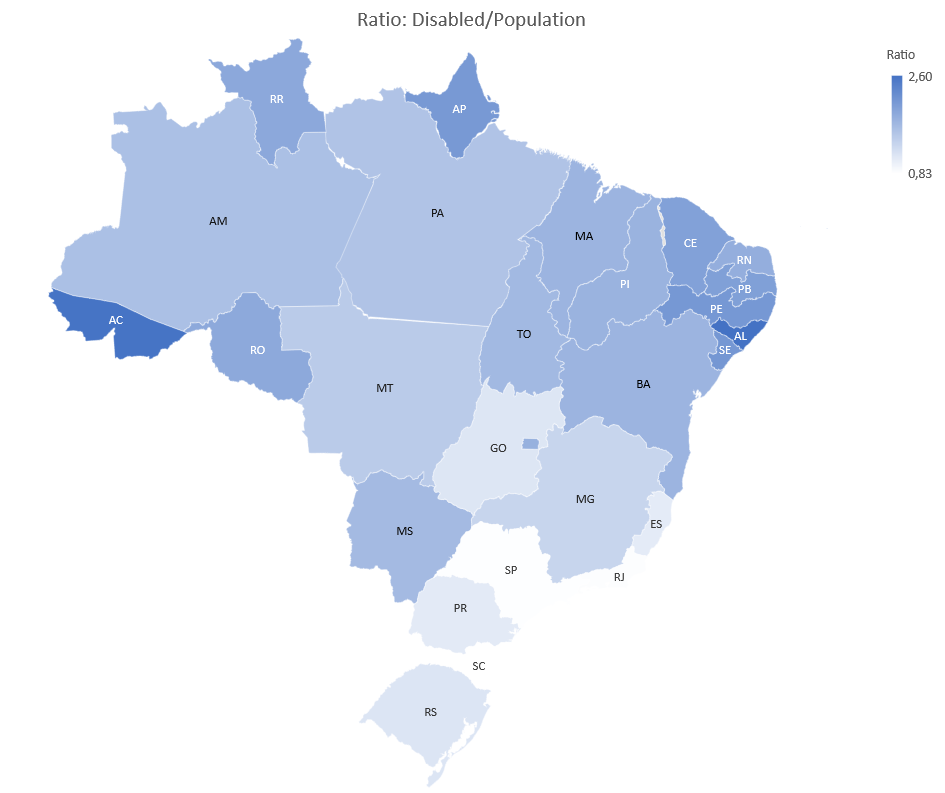}
	\caption{Ratio: Disabled bnf / Population by UF}
	\label{F:map_disabled}
\end{figure}

Figure~\ref{F:map_disabled} displays the Brazilian map using the ratio for Disabled as scale, also to highlight the outliers UF's.

\begin{figure}[H]
	\centering
	\begin{minipage}{.5\textwidth}
		\centering
		\includegraphics[width=.9\linewidth, height=150px]{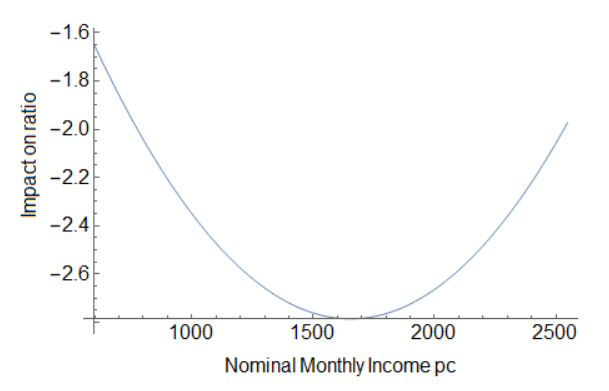}
		\caption{Nominal monthly income on $Y2$}
		\label{F:x2-ydisabled}
	\end{minipage}%
	\begin{minipage}{.5\textwidth}
		\centering
		\includegraphics[width=.9\linewidth, height=150px]{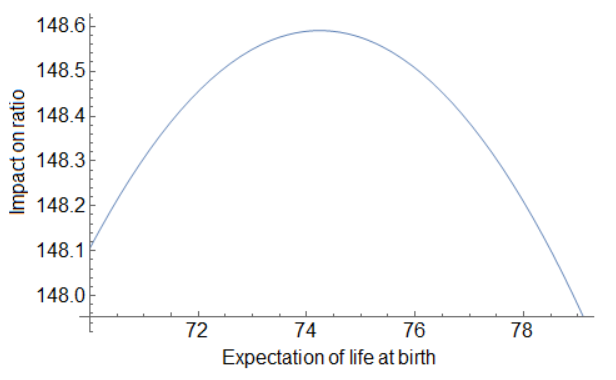}
		\caption{Expectation of life on $Y2$}
		\label{F:x3-ydisabled}
	\end{minipage}
\end{figure}

Figures~\ref{F:x2-ydisabled} and \ref{F:x3-ydisabled} exhibit the impact of both explanatory variables on the ratio. In this model the impact has a similar behaviour as the variables in the first model, considering the Total group.
%
%
%
We can observe in Figures~\ref{F:x2-ydisabled} and \ref{F:x3-ydisabled} that both variables have a linear and a quadratic component.
Until R\$~$1,654$, the Nominal Monthly Income \textit{per capita} decrease the ratio of total beneficiaries by UF. After that turning point, starts to increase. Throughout the range of Income \textit{per capita} the impact is always negative. 
The Life Expectancy at birth has the same behaviour for all three models: Increases the ratio until a turning point and then decreases, in this case the turning point is 74.24 years (the oldest from all models). Also, inside the range of Life Expectancy the impact is always positive.

Some interesting remarks are that although HDI among UFs vary from 0.63 to 0.82, HDI was not statistically significant in any of the chosen models. The same happened to Demographic density, that varies from 2.01 to 444.66 and was not significant either. 
The UF number 22, that represents the DF, was an outlier on all three models. The Nominal Monthly Income \textit{per capita} has a different impact on the models.
The Life Expectancy at birth has a similar behavior, changing only at the age of the turning point: 73.9, 73.6 and 74.24, respectively. 

When we defined the models, in the beginning of Section \ref{S:model}, we had four hypotheses. In all three models the variables HDI and Demographic Density were not statistically significant, for this reason we can not neither accept nor reject any H1 and H4. 
	
Regarding H2, we were testing whether the increase in Nominal Monthly Income \textit{per capita} would decrease the ratio of beneficiaries. On the Total and Disabled models this happens when we start to increase the Nominal Monthly Income, however after the turning point (R\$~$1,429$ and R\$~$1,654$, respectively) it increases and in the case of the Elderly model the ratio increases for the entire range. Therefore, we reject H2.

Finally, concerning H3, we expected to see an increase in ratio when we had an increase in Life Expectancy in the UF’s. In all three cases this is what happens in the beginning, however it is not linear, it is quadratic, which means that after the turning point starts to decrease. So, we do not completely reject our hypothesis. Instead, we change it. 

These results are very interesting due to the fact that they show that you do not capture the heterogeneity among the UF's, in such a way that you could just either increase or decrease one variable and adjust the ratio of beneficiaries. Also, the UF's have similar behaviour, that is,
we observed a geographic pattern among UF's, which lead us to look for clusters. We do that in the section that follows.

\section{Cluster Analysis}
\label{S:cluster}
Cluster analysis is a group of multivariate techniques whose primary purpose is to group objects based
on the characteristics they possess \citep{Hair2014}. It aims to explore data sets to assess whether they can be summarised in terms of a limited number of groups of individuals with some sort of similarities and which are different in some respects from individuals in other clusters \citep{Everitt2011}.

After observing the results in the previous section, we decided to look for clusters, more specifically, if we could have some geographical groups. We analysed hierarchical and non-hierarchical clusters, separately. 
We performed  an analysis based on three different Life Expectancies: At birth, at 60 and 65 years old. Our motivation lies in looking in which cases the Life Expectancy would be lower but because of young people dying (usually as a result of violence).
%

A hierarchical clustering method produces a classification in which small clusters of very similar observations are nested within larger clusters of less closely-related observations.
In a hierarchical classification the data are not partitioned into a particular number of classes or clusters at a single step. Instead, the classification consists of a series of partitions, which may run from a single cluster containing all individuals, to having a single individual per  cluster \citep{Everitt2011}.

We used the Euclidean distance measure to compute the distance matrix and the single method for the cluster agglomeration.
The hierarchical clustering process can be portrayed graphically in several ways. Firstly we present dendrograms. That is a common approach, which represents the clustering process in a treelike graph \citep{Hair2014}.
%
\begin{figure}[h]
	\center
	\includegraphics[width=13cm]{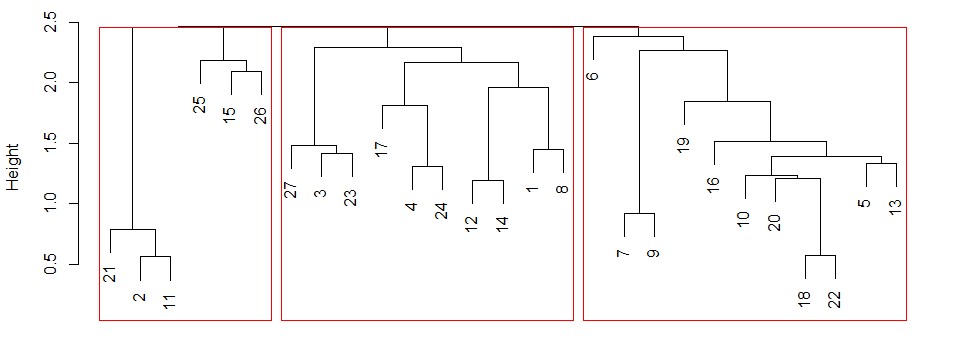}
	\caption{Dendrogram for the expectancy of life at birth}
	\label{F:dend_nascer}
\end{figure}
\begin{figure}[h]
	\center
	\includegraphics[width=13cm]{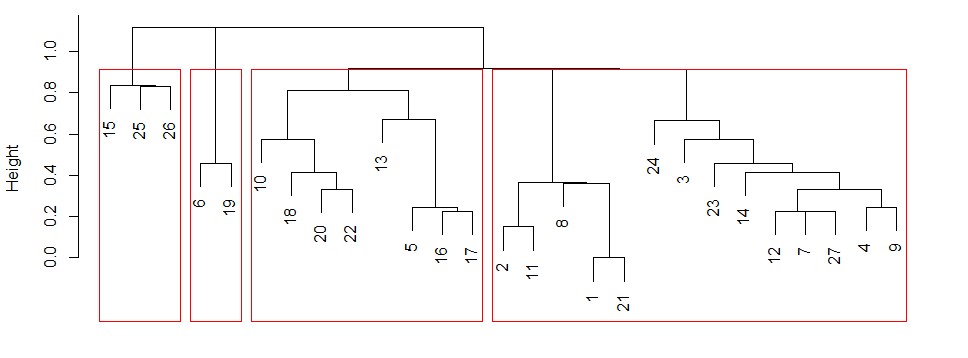}
	\caption{Dendrogram for the expectancy of life for the person that is 60 years old}
	\label{F:dend_60}
\end{figure}
\begin{figure}[h]
	\center
	\includegraphics[width=13cm]{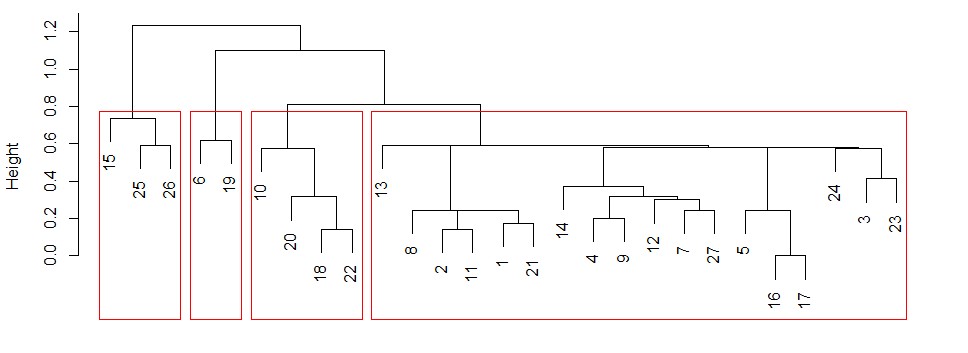}
	\caption{Dendrogram for the expectancy of life  for the person that is 65 years old}
	\label{F:dend_65}
\end{figure}


From the dendrograms we can observe that UF's numbers 21, 2 and 11 that represent \textit{Sergipe, Amazonas} and \textit{Pará} are the farthest left and in a cluster with 15, 25, 26 that represent \textit{Piaui, Rondônia} and \textit{Roraima}. In Figures \ref{F:dend_60} and \ref{F:dend_65} these three (15, 25, 26) are those in the extreme left, while (2, 11 and 21) are in the middle of the dendrogram.
Another fact is that on Figure \ref{F:dend_65} we have a much flatter dendrogram and one of the clusters comprehends 18 out of 27 UF's. 

In Figure \ref{F:dend_nascer}, \textit{Espírito Santo} and \textit{Santa Catarina} (6, 19) are amongst the others, in the biggest cluster. On the regression models these two were outliers, this is shown on the clustering using the Life Expectancy for a person that reaches 60 and 65 years old, respectively. In both cases the two UF's are excluded from the others in a cluster that only contains them.

We can observe in the maps in Figure \ref{F:clusters} the three UF's are clustered together and this cluster consists on people that got to 60 and 65, but present a lower expectancy. These three UF's are \textit{Piauí} in the North-East region and \textit{Rondônia} and \textit{Roraima} from the North region. From these maps we can see a high geographic correlation on the homogeneity amongst the clusters.

%
In contrast to the  hierarchical methods, non-hierarchical procedures do not involve the treelike construction process. Instead, they assign objects into clusters once the number of clusters is specified  \citep{Hair2014}.
The non hierarchical clustering algorithms are used mainly for extracting compact clusters by using global knowledge about the data structure \citep{Dzwinel2004}.

In the case of people that are 60 years old, the UF's 6 and 19 are in a separated cluster both in the hierarchical and non-hierarchical analysis. 
\textit{Rio Grande do Norte} is the only UF from the upper part to be clustered (in grey) with the UF's from the lower part of Brazil.

On the Life Expectancy at birth,  regions \textit{South, Center West} and \textit{South East} constitute one sole cluster, both when clustering hierarchically or non hierarchically. 
The highest difference between these two structures is in the third case, people that are 65 years old.   
When you consider people that already lived until 65, in the \textit{South-East} region \textit{Rio de Janeiro} is clustered along with states from the upper part in Brazil. This happens in the hierarchical and non hierarchical cases.

We showed the existence of {high heterogeneity, for some chosen economic variables, on the population of beneficiaries among Federal Units (UF's), specially among the five regions in Brazil }(North, North-east, Centre-west, South-east and South), which has an obvious impact on the beneficiaries of the programme. 

Next we go much further, using a different database, we want at first to calculate the amount that INSS spends with the beneficiaries, estimate values \textit{per capita}, then the weight of each UF and see if this high heterogeneity is also reflected on the {benefit amounts}. For the estimate calculation, we use a more comprehensive database that includes every Brazilian that started receiving a pension or a benefit from the INSS in the period from 2nd of January 2018 to 6th of April 2018. Using this database we focus on the beneficiaries of this specific programme. In the first database we had the same population but with fewer information and in this second despite only using a sample of data, we have more specific information on every individual.

\section{Expected Discounted Benefit}
\label{S:edb}
In this section we calculate an estimate of the Expected Discounted Benefit (briefly, EDB) for the beneficiaries of our sample in the elderly group. It is indeed an estimate of the expected present value of the future monthly benefits, at some discount rate. As aforementioned (end of previous section) we use the Life Expectancy for an individual that is already 65 years old, since this is one of the criteria to be eligible for the programme. We chose an annual discount rate of 6\%, this is the official return rate for Brazilian treasure bonds in 2019. Since the benefit is a monthly payment we use the equivalent monthly rate of $0.486755\%$. It is clear that there is a greater heterogeneity on life expectancies among UF's as well as between genders. Since the poorer states have a lower LE, it is clear that most beneficiaries of the programme live in the wealthier states, despite the fact {that it can be argued that most of public money come from those richer}. Although, we could counteract that public social policies should target the poorer regions. Knowing this, we can pose the following question: Is there a way to narrow differences among these sub groups? We can do changes on benefits {(often not well viewed)} or, better, change retirement ages {without doing any changes} to the benefits. Both approaches have \textit{pros and cons}, we will come to this later. We observe for instance that in some poorer UF's life expectancies are much lower, much lower in some cases, resulting in many people not taking the benefit where it would be needed more, in the sense that there are poorer people, relatively speaking. Our starting tool is the calculation of the Expected Discounted Benefit.

To explain the calculation technique used let's put some mathematics, although simple. First define locally some quantities, in order to set clear the calculation of the necessary Age Adjusting Factors (briefly AAF). We show the calculation separating by gender, although the second proposal does not separate but calculation method is likewise.

Let $b$ be the minimum monthly wage which equals the monthly benefit. Let $C(i,j,k,l)$ be an estimate for the present value of the benefit cost by UF $i$, sex $j$, in time $k$, for individual $l$. The variable time here is measured in months after some reference age. For simplification we consider that if a beneficiary does not live a full month the benefit is not paid, benefits are paid at the end of each month ({to be precise in our case it is on the 28th or 29th}). Let $r$ be the monthly equivalent discount rate. {Here we }have $r=0.00486755\simeq (1.06)^{(1/12)}-1$. Also, let $n_i$ be the number of beneficiaries in UF $i$, $LE_{ij}$ is the Life Expectancy in UF $i$ for sex $j$ ($i=1,\dots,27$, $j=1,2$). It's clear we are using 54 Life Expectancies. Define $LT_{lij}$ as the Expected Lifetime of individual $l$ in UF $i$ for sex $j$, in exceeding months after a reference date, we use the date of 06/04/2018, the last day of granting benefit in our sample. This way we garantee that all individuals in our sample are already receiving a benefit. For instance, if Life Expectancy is 69 exactly for some for individual $l$ in state $i$ and  sex $j$, with granting age 65 years and  2 months (in 06/04/2018), then $LT_{lij}= 46$ months.  Let $n=\sum_{i=1}^{27}=n_j= 40,372$ be the number of beneficiaries in our sample. In another way, let's consider also $ n_i=m_{i1}+m_{i2} $, separating by gender (subscript 1 stands for male, female otherwise), where $m_{ij}$ is the number of people of sex $j$ in UF $i$, $ i=1,2,\dots,27 $ and $ j=1,2 $. Following a similar notation reasoning we set $m_j=\sum_{i=1}^{27}m_{ij}$ as the number of people of sex $j$ in the whole country. Then $n=m_1+m_2=17,351+23,021=40,372$, by order male and female.

In our sample, for individual $m$ from UF $i$ with sex $j$, we have that 
\begin{eqnarray*}
	C(i,j,k,l)& = &\frac{b}{(1+r)^k} \\
	{C(i,j,l)} &= &\sum^{LT_{lij}}_{k=1}\frac{b}{(1+r)^k}=b\, a_{\lcroof{LT_{lij}}\,r}\\
	C(i,j) &=&  \sum^{n_i}_{l=1} {C(i,j,l)}\,.
\end{eqnarray*}
It is clear that $C(i,j,l)$ is the present value of all benefits received by individual $l$ until $LT_{lij}$, and $C(i,j)$ is the {present value of  all benefits} in UF $i$ for sex $j$. $a_{\lcroof{n}\,r}$ is the standard formula of the temporary unit payment, with term $n$ and discount rate $r$, financial annuity.  For simplification we consider that every individual is receiving  in 6/4/2018 the full amount of the monthly benefit. We use the $LT_{lij}$ for all individuals of UF $i$, it is a clear simplification because {there are} no mortality tables available for each UF, only the national ones.

Then for the whole of Brazil, we have 
\begin{eqnarray*}
	C(j) &=&  \sum^{27}_{i=1} {C(i,j)}\\
	C &=&  \sum^{2}_{j=1} {C(j)}\,.
\end{eqnarray*}

The Expected Discounted Benefit for the our sample is  $\text{\euro}\,1,105,411,797.02$. When considering Life Expectancy at birth the value estimate equals $\text{\euro} 714,044,109.82$, this big difference is brought by the LE's. The real amount that the INSS will spend with the beneficiaries in our sample is a value between these two and closer to the first one. We used the exchange rate from the day 04/04/2019, making \euro 1 corresponding to R\$~4.35, therefore we have for Life Expectancy at 65 and at birth the amounts R\$~$4,808,541,317.02$ and R\$~$3,106,091,877.73$, respectively. From now on all values are shown in Euros. 

We are interested in calculating the necessary Expected Discounted Benefit for a beneficiary in the programme, that starts receiving the benefit now. From this amount we could estimate the necessary amount for the entire population of beneficiaries. However, we do not require this latter value to achieve our goal, due to the fact that we want to analyse the Expected Discounted Benefit estimate per capita by UF, looking for heterogeneity among UF's.

\begin{figure}[h]
	\center
	\includegraphics[width=11.8cm]{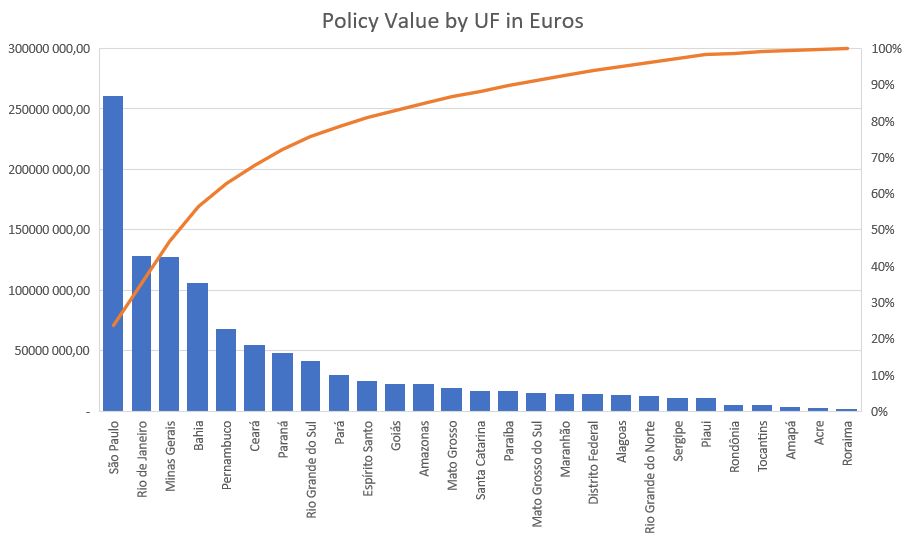}
	\caption{Expected Discounted Benefit (EDB) by UF's in Euros}
	\label{F:reserve}
\end{figure}

Figure \ref{F:reserve} shows in blue bars the amount corresponding to the Expected Discounted Benefit by UF (absolute values), while in orange we show the cumulative amount {(in \%)}. {Under the bars} are the corresponding names of the UF's. We can draw attention to \textit{São Paulo} (it is the wealthiest UF) that corresponds to 23.61\% of the total amount while \textit{Roraima} (the UF with smallest population and the lowest GDP) measures up to only 0.19\%. We also need to take into consideration the fact that the population of these UF's is of 41,262,199 and 450,479 people, being the most and {the least} populated Federal Units in Brazil respectively. Therefore, obviously we can not use the Expected Discounted Benefit by itself, so we will consider the Expected Discounted Benefit per capita. We can already spot heterogeneity among UF's. 

\begin{figure}[h]
	\center
	\includegraphics[width=12cm]{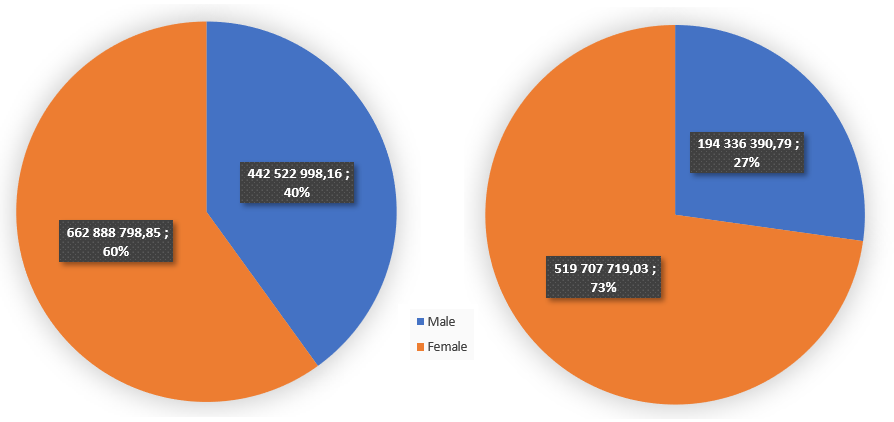}
	\caption{Expected Discounted Benefit (EDB) in Euros: LE at Birth and at 65}
	\label{F:reserve-pie}
\end{figure}

\begin{figure}[h]
	\center
	\includegraphics[width=10cm]{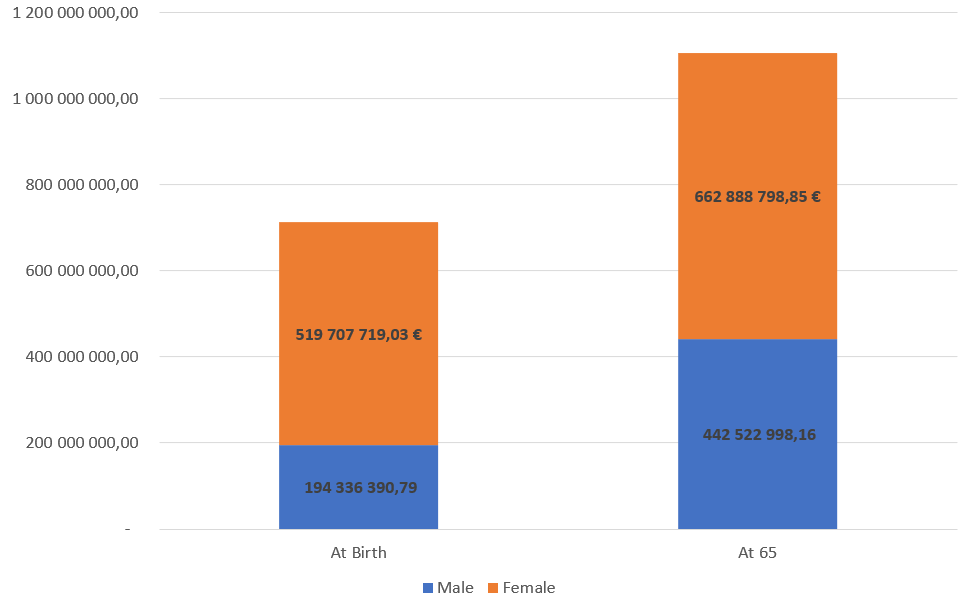}
	\caption{Expected Discounted Benefit (EDB) in Euros: LE at Birth and at 65}
	\label{F:reserve-col}
\end{figure}

In Figures \ref{F:reserve-pie} and \ref{F:reserve-col} we present the distribution of Expected Discounted Benefit separating by sex. 
It is important to highlight two facts. The first is that there is a huge difference between genders. In Figure \ref{F:reserve-pie} we can observe that when we consider LE at birth, the male beneficiaries represent only 27\% and when considering LE at 65 it represents 40\%. This is due to the fact that in Brazil there is a large disparity in Life Expectancy between genders. For instance, in the UF \textit{Ceará} this disparity is of 8.28 years on LE at birth. The disparity varies from 4.99 to 8.28 with an average of 6.8 years for LE at birth. For LE at 65 it varies from 1.9 to 4 and the average is 3.03. That is why on the case of LE at 65 the female group represents 60\% of total, it portrays the decrease of the difference between genders.

The second fact is the disparity that we can observe from Figure~\ref{F:reserve-col} between the Expected Discounted Benefit calculated with different Life Expectancies. The reason for this is the fact that in Brazil a significant part of the population dies before 65.

We aim to analyse the Expected Discounted Benefit by Federal Unit and subsequently try to make the system more homogeneous. Not rejecting alternatives, a simple but effective method we propose {is creating} a kind of \textit{actuarial} correcting factor to take into consideration the Life Expectancy disparities among Federal Units. We also separate by gender in one of the proposals. This is done in a way that considers that every beneficiary in Brazil presents the same Expected Discounted Benefit estimate. Although we considered the LE at 65 to see how much the government will spend by UF, for the purpose of policy making we will consider LE at birth only. This is because we need to take into account that there are UF's where a greater part of the population will not reach 65. This coincides with the poorer UF's, those were the UF's that were supposed to need more support from this programme.

\begin{figure}[h]
	\center
	\includegraphics[width=14cm]{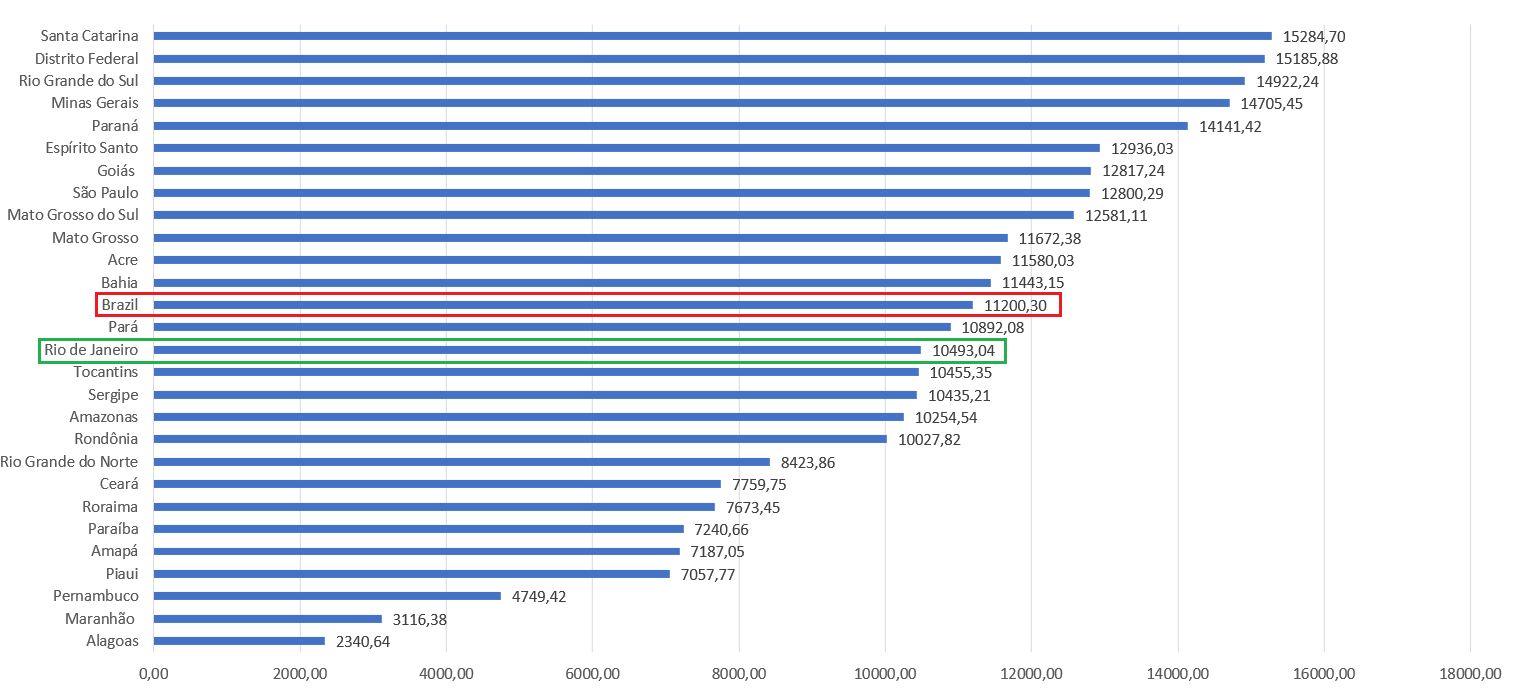}
	\caption{Expected Discounted Benefit (EDB) - Male Group}
	\label{F:reserve-male}
\end{figure}

\begin{figure}[h]
	\center
	\vspace{-0.3cm}
	\includegraphics[width=10cm]{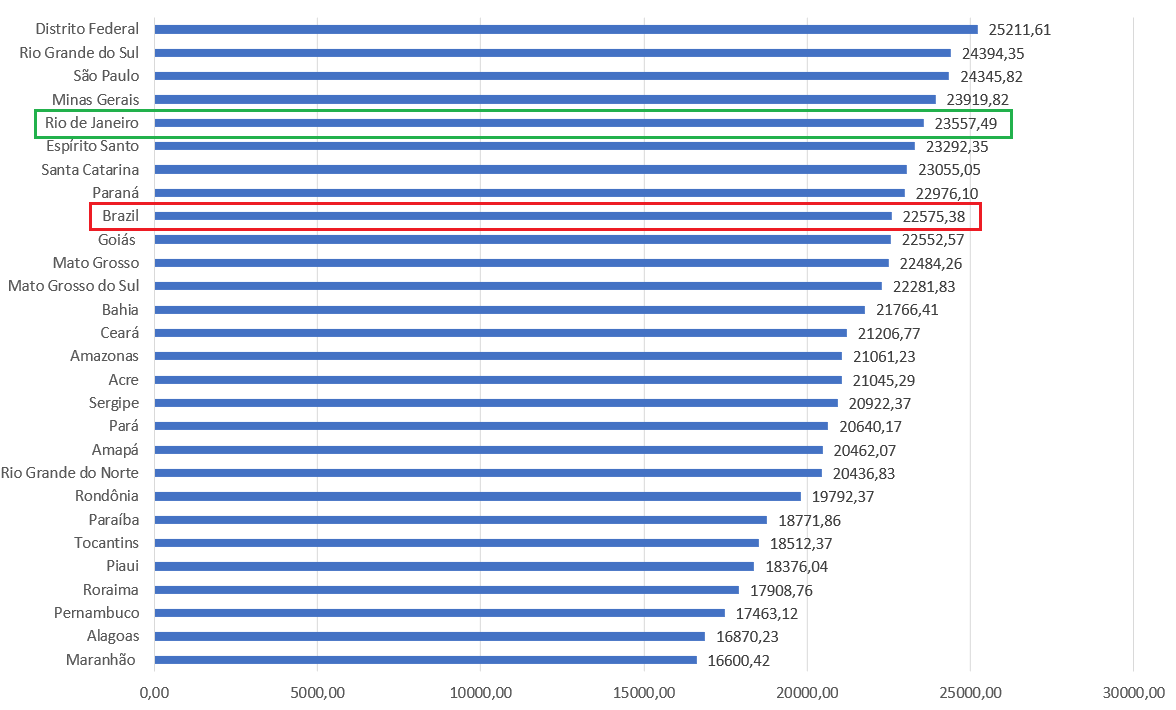}
	\caption{Expected Discounted Benefit (EDB) - Female Group}
	\label{F:reserve-female}
\end{figure}

\begin{figure}[h]
	\center
	\vspace{-0.3cm}
	\includegraphics[width=10cm]{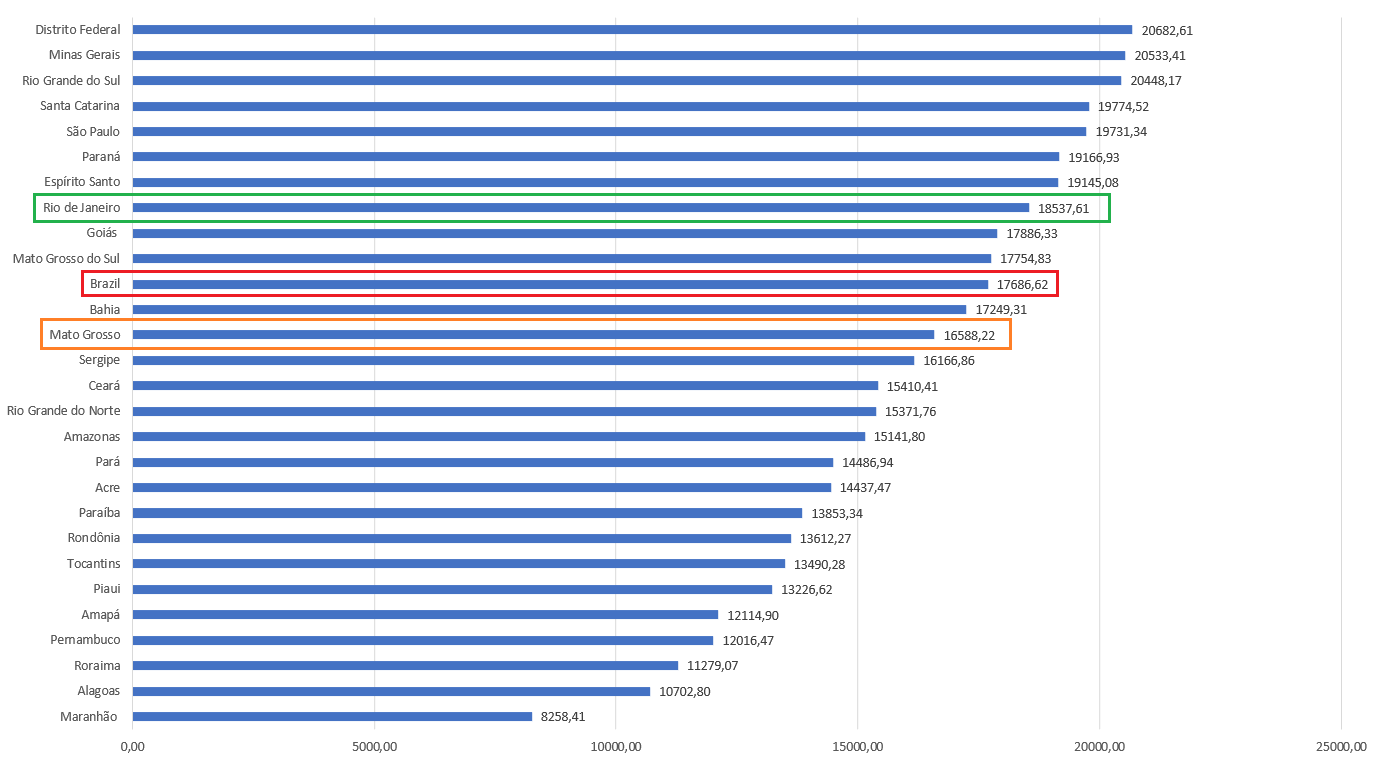}
	\caption{Expected Discounted Benefit (EDB) \textit{per capita}}
	\label{F:reservepc}
\end{figure}

Figures \ref{F:reserve-male}-\ref{F:reservepc} show the Expected Discounted Benefit per capita, first for the male group, then the female group and finally the total. On these figures we will represent inside the red frame is the average of the beneficiaries in our sample, Brazilian's average estimate.

In Figure \ref{F:reserve-male} we exhibit {the EDB by UF in descending order of amount}. {Amounts vary from}~\euro $2,340.64$ to \euro $15,284.70$, so we can clearly spot the heterogeneity among UF's, with the exception of the UF \textit{Rio de Janeiro}, marked green, \textit{Acre} (North) and \textit{Bahia} (Northeast), the UF's above the average are from the regions: South, Southeast and Centre-West and those below the average are from the North and Northeast region. 

In the case of the female group, we can observe in Figure \ref{F:reserve-female} that the UF that presents the lowest Expected Discounted Benefit per capita is \textit{Maranhão}, with the amount of \euro $16,600.42$. This figure is already higher than the highest Expected Discounted Benefit per capita from the male group, that was \textit{Santa Catarina}, with \euro $15,284.70$. Therefore, there is no intersection between these two intervals, being the latter from \euro $16,600.42$ to \euro $25,211.61$.

Another aspect that we want to highlight is the fact that \textit{Rio de Janeiro}, which is below average in the male group, is located here as one of the top five UF's (violence may be the cause). Subsequently, it dragged the average with it and in Figure \ref{F:reserve-female} above average we have the Federal District and UF's from the South and {South-east regions}. Below average, once more, the North and North-east regions and now, also the Centre-West region that was previously above average (on the male group, Figure~\ref{F:newage-male}).

Figure \ref{F:reservepc} presents the whole group, the {interval goes} from \euro$8,258.41$ to \euro~$20,682.61$. \textit{Rio de Janeiro} is once more above average and when we compare to the average, only \textit{Mato Grosso}, in orange, an UF from the Centre-West region is below. The rest of the UF's from the Centre-West region and the UF's from South and Southeast region are above it and once more all of the UF's from the North and Northeast region remain below average.

After analysing these figures with the Expected Discounted Benefit per capita we could attest the existence of a clear heterogeneity. In Figure \ref{F:reserve-male} we have the beneficiaries with the highest Expected Discounted Benefit per capita costing over 16 times more than the lowest. In the female group this difference drops down clearly but the EDB per capita is still over 50\% higher when comparing UF's. For the total group this gap is of almost three times the value of the lowest EDB. Therefore, we will propose two different factors to try to achieve a fairer system, named as \textit{Age Adjusting Factors}. We will calculate 54 and 27 different factors, as we will consider this factor depending on both UF and sex and just depending on {UF, respectively. This} is going to be discussed in Section~\ref{S:af}.
\section{Proposing Age Adjusting Factors on the Benefit Age}
\label{S:af}
We present two proposals for Age Adjusting Factors, in the subsections below. These factors were calculated so that we get the same Expected Discounted Benefit per capita for the beneficiaries in Brazil, independently of the UF. We do not change the value of the benefit (minimum wage), instead, we change the minimum {age of application for this} micro programme. The resulting age is $AAF \times 65$, in each proposal we denote the AAF's as $AAF_1(x,y)$ and $ AAF_2(x) $, respectively. The first depending on UF and sex and the second only on UF, denoted by $x$ and $y$.

We set $\bar{C}(j)= C(j)/m_j$, $j=1,2$ as the (discounted) average cost of benefit per gender, independent of UF, $m_j = \{17,351;\,23,021\}$ for male and female respectively.
%

Now,
define $D(i,j)$ to be the average benefit difference to the \textit{national} average for UF $j$ and sex $i$, such that
\begin{eqnarray*}
	D(i,j)& =& \bar{C}(j) - \bar{C}(i,j) \\
	\bar{C}(i,j)&=& {C}(i,j)/m_{ij} \, \text{ } i,j=1,2,\dots 27;\,j=1,2\,,
\end{eqnarray*}
and consider
\begin{eqnarray}
|D(i,j)| &= &b\, a_{\lcroof{z_{ij} }\,r}\,, 
\label{e:dij}
\end{eqnarray}
%
%
where $z_{ij} =|w_{ij}|$ represents the time (in monthly periods) necessary for the annuity on the righthand side of~\eqref{e:dij} to equal $|D(i,j)|$ ($w_{ij}$ can be negative and only depends on UF and sex). If the corresponding $D(i,j)$ is negative then so will be $w_{ij}$ (positive otherwise, if it is zero, there's no need to change LT). Now, if we consider $MLT_{lij} = LT_{lij} + w_{ij} $ we will get the same expected discounted benefits.  MLT is the Modified LT in order to get equal cost per capita in all UF's (in the first proposal below we consider also equal cost to all UF's and genders). 

The Age Adjusting Factor, in Proposal~1, following the above method comes
\begin{eqnarray*}
	MLT_{ijl} &= &LT_{ijl} + w_{ij} \\
	AAF_{1}(i,j)& = &\frac{65 + w_{ij}}{65}\\
	NA_{ij} & = & AAF_{1}(i,j) \times 65 \,,
\end{eqnarray*}
where $NA_{ij}$ stands for new starting age for the benefits. For the $AAF_{2}(i)$ case, the calculation is analogous, the sole difference comes from using here $\bar{C}=\sum_{j=1}^{27}m_j/n$ instead of $\bar{C}(j)$, $j=1,2$ in each case, since we do not {separate} by gender.
\subsection{First Proposal}
Our first proposal separates the group of beneficiaries between male and female. We propose new ages for each UF and gender, resulting in 54 different factors, denoted as  $AAF_1(x,y)$, where $x$ and $y$ stand for UF and Sex, respectively. 

\begin{figure}[h]
	\center
	\vspace{-0.3cm}
	\includegraphics[width=7.2cm]{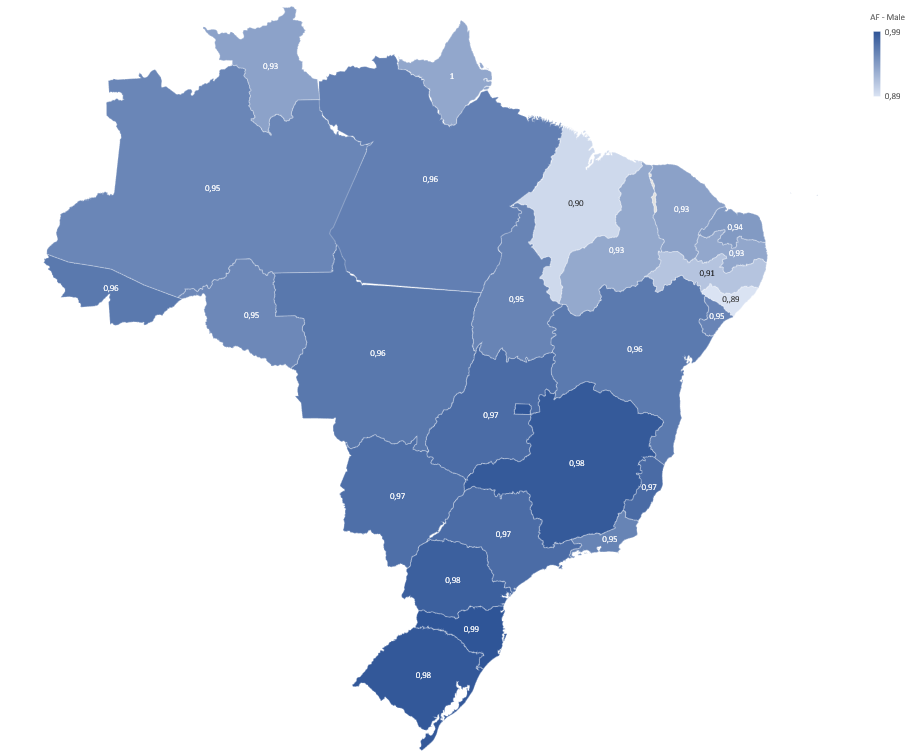}
	\caption{First Age Adjusting Factor ($AAF_1$) - Male}
	\label{F:af-male}
\end{figure}

\begin{figure}[h]
	\center
	\vspace{-0.3cm}
	\includegraphics[width=7.2cm]{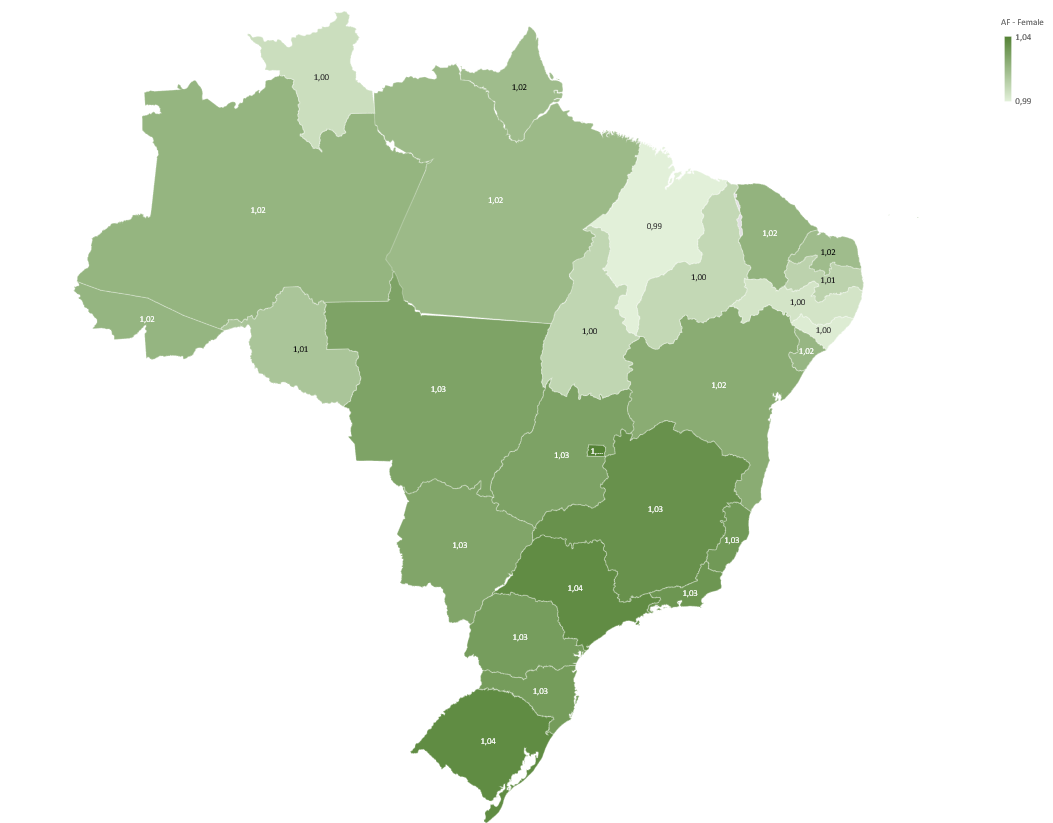}
	\caption{First Age Adjusting Factor ($AAF_1$) - Female}
	\label{F:af-female}
\end{figure}

In Figures \ref{F:af-male} and \ref{F:af-female} we present the age adjusting factors: {In the graph we show in blue the male group, and in green the female group}.  We can easily notice that both blue and green are lighter on the upper part from the map and darker on the bottom one. In Figure \ref{F:af-male} the age adjusting {factors vary} from 0.89 to 0.98.

For the female group we can observe in Figure \ref{F:af-female} that the age adjusting factor varies from 0.99 to 1.04, which means that while for the male group the required age would decrease for all of them, for the female group it would decrease for part of them but for most of the group it would increase. 

In the calculation we set the starting national expected discounted benefit figure as irrespective of gender. Once more it is shown there is a serious gender problem, as all male newly calculated retiring ages are lower than 65 and almost all females' are higher. 

\begin{figure}[h]
	\center
	\includegraphics[width=10cm]{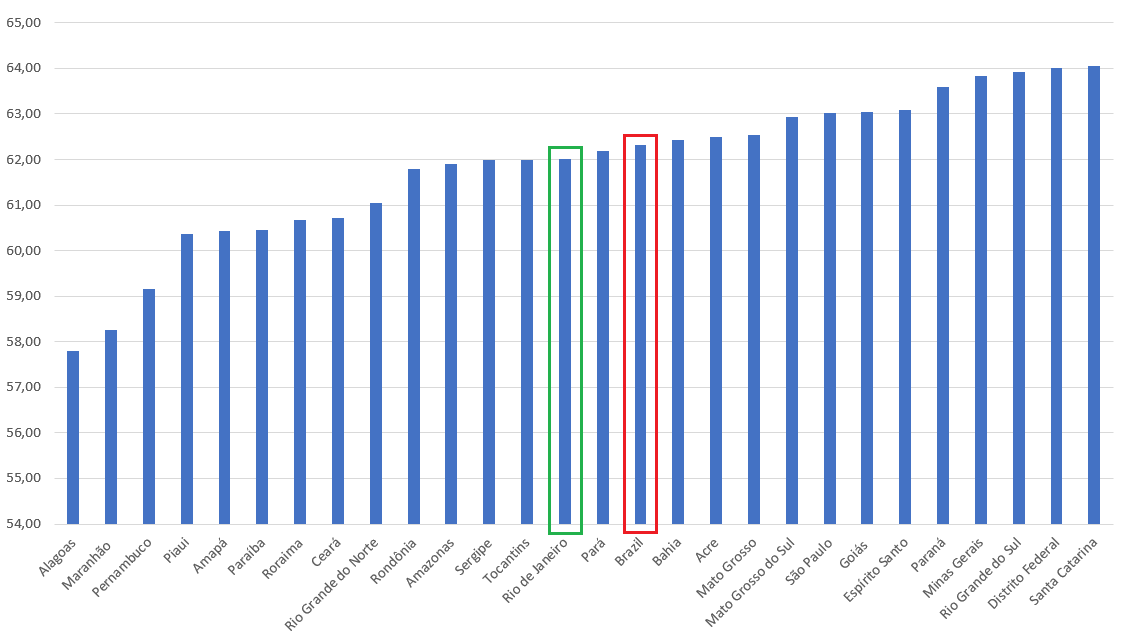}
	\caption{New age for the Male group}
	\label{F:newage-male}
\end{figure}

\begin{figure}[h]
	\center
	\includegraphics[width=9.8cm]{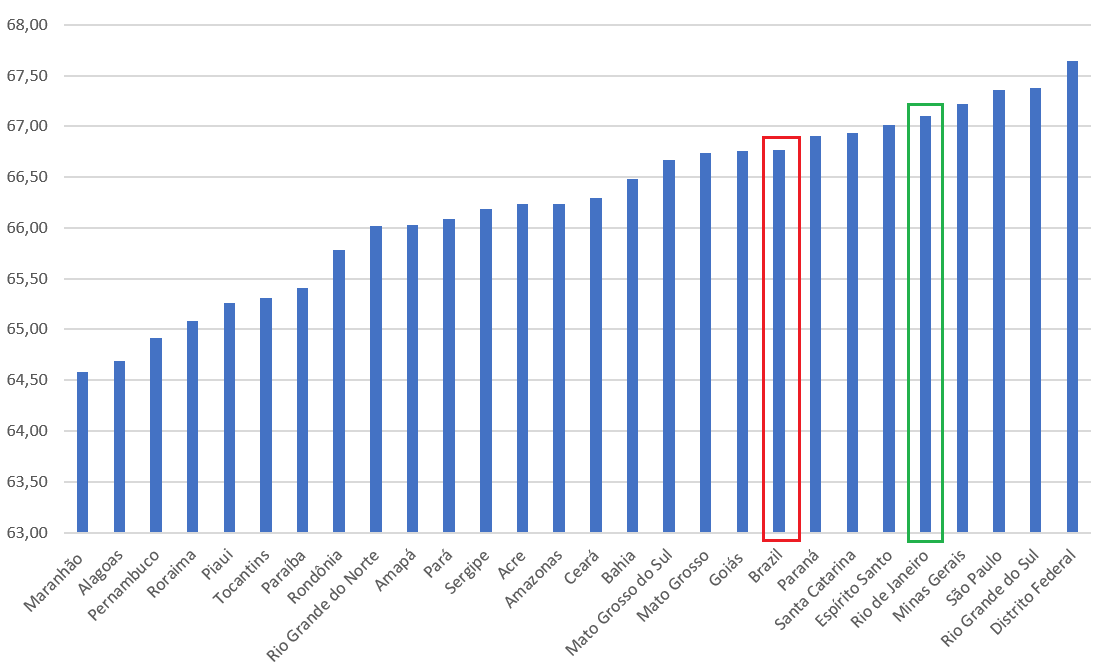}
	\caption{New age for the Female group}
	\label{F:newage-female}
\end{figure}

Figures \ref{F:newage-male} and \ref{F:newage-female} show the new ages in another way, by bar graphs. For males, since the age adjusting factors vary from 0.89 to 0.98, the new age for the male group will be less than 65 for all UF's, varying from 57.80 to 64.06. Similarly to Figure \ref{F:reserve-male}, we have highlighted in red Brazil (the average) and in green \textit{Rio de Janeiro}. 

In Figure \ref{F:newage-female} we present the proposed new ages for the female group that vary from 64.58 to 67.64. As mentioned in Section~\ref{S:edb} there is no intersection between the ages in these two groups. 

\subsection{Second Proposal}
\begin{figure}[h]
	\center
	\includegraphics[width=11.8cm]{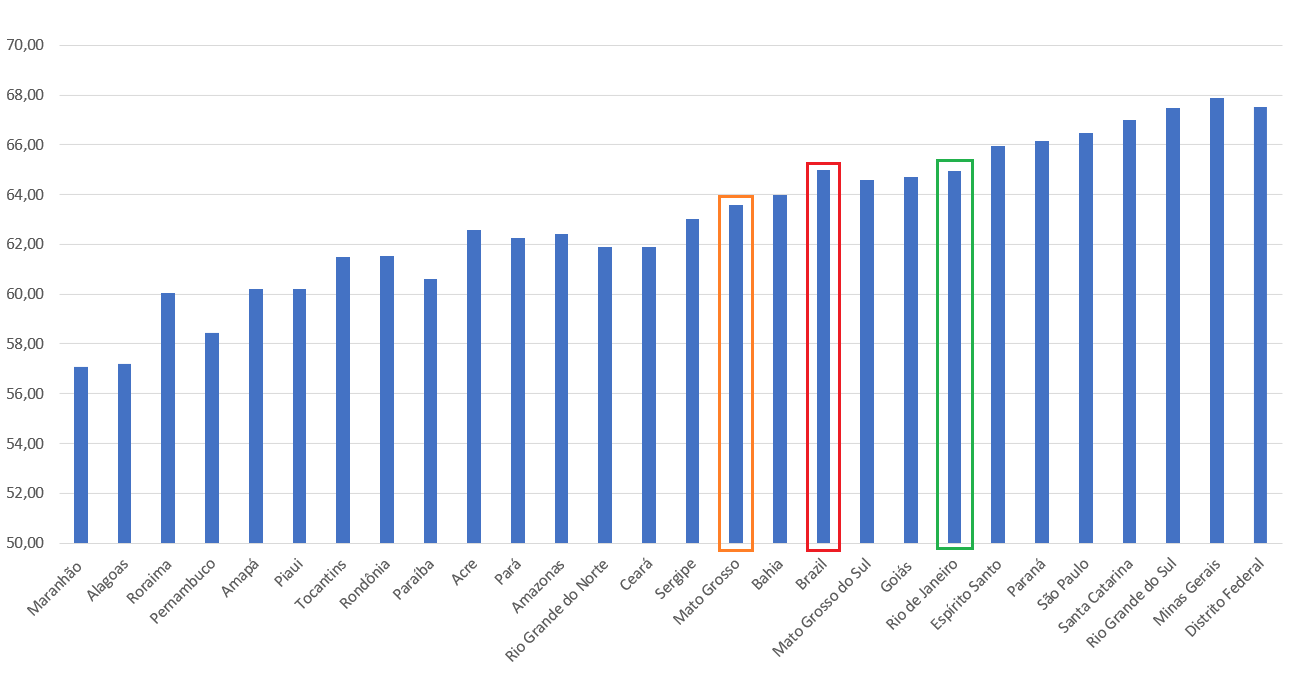}
	\caption{Our Second Proposal: New age}
	\label{F:newage}
\end{figure}

\begin{figure}[h]
	\center
	\includegraphics[width=8cm]{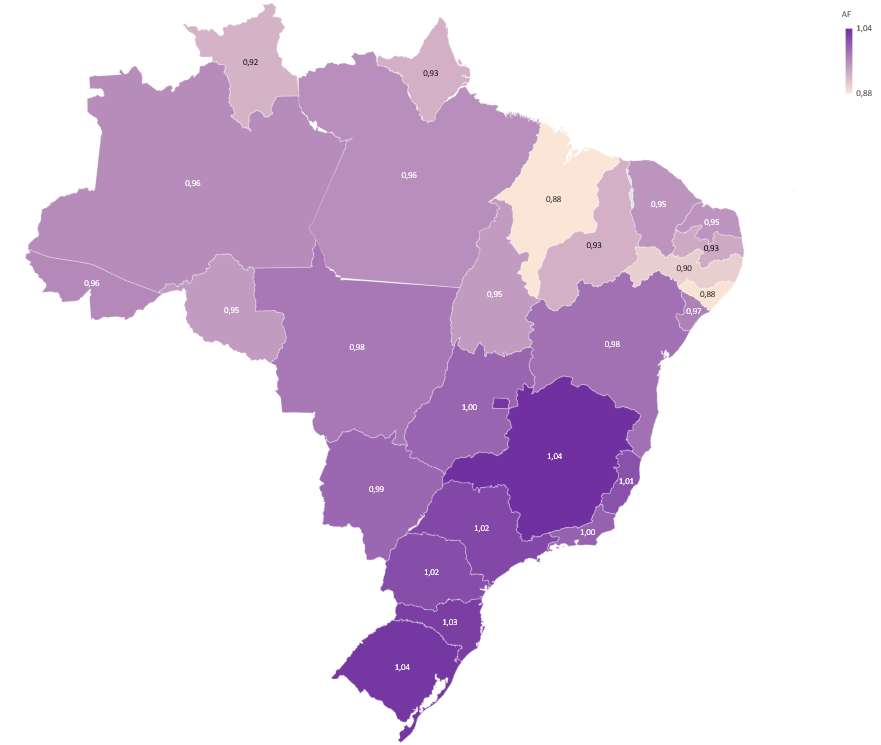}
	\caption{Second Age Adjusting Factor ($AAF_2$)}
	\label{F:af}
\end{figure}

Our second proposal does not separate groups by gender. Recalling what we wrote in Section \ref{S:edb} when describing Figure \ref{F:reserve-pie}, for those people that reach 65 years old, the gap in Life Expectancy between genders decreases from 4.99-8.28 to 1.9-4.  
Similarly to the prior subsection we show in Figures \ref{F:newage} and \ref{F:af} the proposed \textit{new ages} and the adjustment factors in the Brazilian map.

In Figure \ref{F:newage} we present the new ages for the UF's in the same order as in Figure \ref{F:reservepc}, so that we can clearly see the differences, they vary from 57.07 to 67.89. The UF's AAF present a behaviour in a way that \textit{Rio de Janeiro} is the only UF from the Southeast region below average. The remaining UF's from Southeast, the Federal District and South region are above average, and the UF's from Northeast, North and Centre-West, are below average.

In Figure \ref{F:af} we show the adjusting factors, we recall that the factor here is only a function of UF residence, and its range goes from 0.878 to 1.044. Once more we can observe a darker colour for UF's from the South and South-east regions, the lighter colours appear in the upper part of the Brazilian map only.
\section{The Reform}
\label{S:reform}
At present there is a proposed reform under way targeting the entire Social Security System in Brazil. At first the proposal of the government was the increase of the minimum age to 70 years old for applying for this benefit.  In addition to this new criteria the household could not possess any assets above the value of R\$$\,98,000.00$, approximately \euro$\,22,528.74$. According to \cite{Fipezap2019} the average price of the squared meter in February 2019 was R\$~$\,7,189$ (\euro$\,1,652.64$) among the 50 cities monitored. Considering this average figure, the limiting upper value would mean a house of $13.63$ square meters.

So the benefit would work in the following way: For each individual, The household could own assets under \euro$\,22,528.74$ and income per capita of less than a quarter of the minimum monthly Brazilian wage. When the individual reaches the age of 60  he would be able to receive a fixed benefit of R\$~$\,400.00$. However, this value is not indexed on anything. Then, and only then, when reaching 70 years old the beneficiary could receive the minimum monthly Brazilian wage \citep{PoderExecutivoBrasileiro2019}. If Life Expectancy is already lower than 65 in many UF's we can imagine how much will be the percentage of population eligible for the programme, living long enough to be a beneficiary.

\begin{figure}[h]
	\center
	\vspace{-0.3cm}
	\includegraphics[width=8cm]{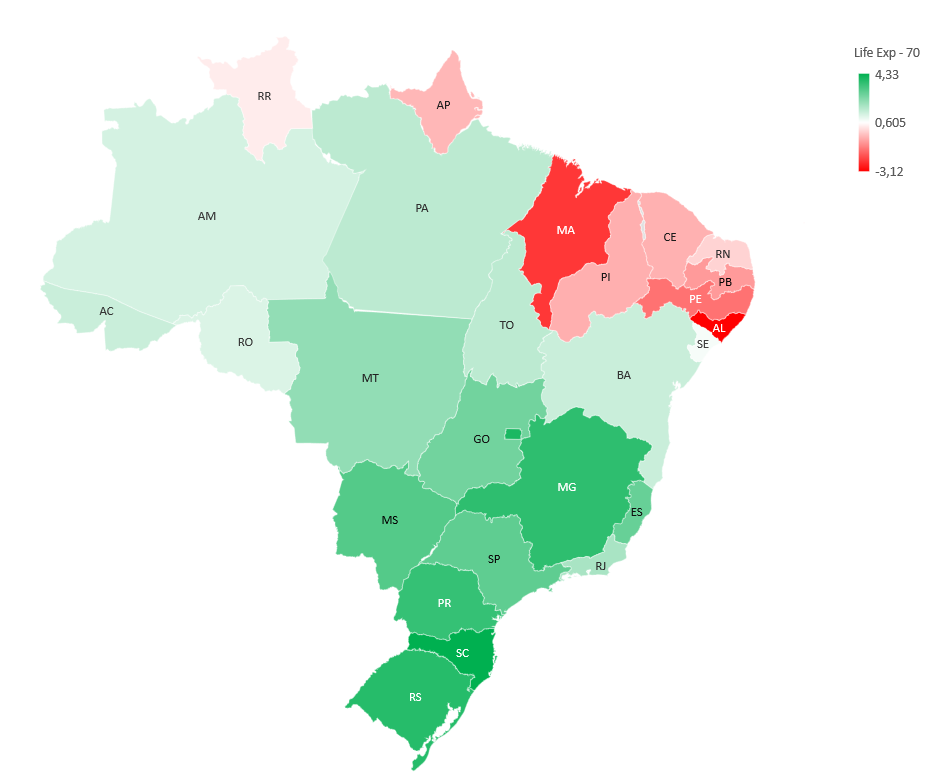}
	\caption{Life expectancy at Birth (male) - 70 years}
	\label{F:expdevida-70}
\end{figure}

Figure \ref{F:expdevida-70} shows the discrepancy between the male Life Expectancy at birth for each Federal Unit and the new proposed age: 70 years old for the entire population. The {interval goes from} $ -3.12 $  to $ 4.33 $ (negative values are in red and positive in green).
This reform is still to be voted and settled, specialists asked to remove the new age criteria from the reform, keeping only the quarter of the minimum wage and the sum of the assets not being greater then \euro$\,22,528.74$.

\section{Some concluding remarks}
\label{S:concl}
%
%
In this paper we aimed to analyse quantitatively the microinsurance programme in the Public Pension System in Brazil. We did so looking from two different angles. Firstly we saw the quantitative impact of four variables: Human Development Index, Nominal Monthly Income \textit{per capita}, Life Expectancy at birth and Demographic Density in the amount of recipients of the social support benefit. We set four hypotheses regarding four economic factors. After that, we calculated the impact that the beneficiaries have on the system and created two Age Adjusting Factors to make the system less uneven. The use of this kind of analysis with such  data was pioneering, we found a lack in the study of this particular programme.

Throughout our study
we used ratios instead of the nominal amounts because the population size among Brazilian Federal Units is quite heterogeneous. In the three models used, we found that from the variables used, Human Development Index and Demographic Density were not statistically significant, which led us to neither accept nor reject the hypotheses H1 and H4, in Section \ref{S:model}.
On the other hand, two had a statistically significant impact on the ratios.
These variables are Life Expectancy and Nominal Monthly Income. The Nominal Monthly Income has a linear and quadratic component in the first and the third model, and only the quadratic one on the second. That agrees with the inequality not just among UF's but inside each. The Life Expectancy has a similar behaviour in all three models. This made us to reject H2 and change our hypothesis H3.

After analysing the regression results we had the idea of looking for clusters. To do so we used the Life Expectancy at birth, at 60 and 65 years old (the Life Expectancy presented a similar behaviour in all three models). In the maps we can clearly attest some geographic clusters and the main distinction between the hierarchical and non-hierarchical clustering lies on the third case, the Life Expectancy at 65 years old.

With the results from this paper we can establish a high heterogeneity among UF's and especially among regions in Brazil. Then, we computed the impact of the beneficiaries on the Programme, looking further for heterogeneity. We mean, computed their Policy Value and checked that there is also heterogeneity. 
%
%
We used two different Life Expectancies (LE): (1) At 65, to calculate the Expected Discounted Benefit so that the Government knows how much they expect to spend; and (2) At birth, for the Age Adjusting Factor to take into consideration those people who will not make to 65 - Public Policy.

For the sample analysed, the Expected Discounted Benefit { equals $\text{\euro} 714,044,109.82$ or $\text{\euro}\, 1,105,411,797.02$ for Life Expectancies at birth and 65, respectively.} 
The female group requires 60\% of the total amount of EDB if we consider the LE at 65 and 73\% for LE at birth.

Considering the male group, with the exception of \textit{Rio de Janeiro}, states on top are from South, South-east and Centre-West and then from North and North-east.
In the Female group, above average {UF's} are DF and from South, South-east and the bottom 16 are the UF's from North and North-east. For the entire  group, {with the exception of \textit{Mato Grosso}}, above average {UF's} are from South, South-east and Centre-west as well as below North and North-east.

We proposed two Age Adjusting Factors to balance the EDB \textit{pc} by UF, one by gender and the other a general one. With the proposed reform of the Public Pension System, in the states that need the most, most of the target population will not live long enough to receive  any benefit.
Also,
it's necessary to ask the Government to better protect the elderly living in these states, since they are less likely to cope during a crisis. The proposed reform will exclude (on average) states from NE.

Two more important remarks deserve our attention: On one hand, our proposals allow that the benefit amounts are maintained equal for every beneficiary, irrespective of Federal Unit; On the other hand, if a proposal like ours is put in place without any restriction, there is always the danger of people in need of the programme moving to a different state to get an earlier benefit. This {could turn the} Federal Units even more unequal than before. 

\section*{Acknowledgements} 
{Authors gratefully acknowledge the financial support from FCT/MCTES - Fundação para a Ciência e a Tecnologia (Portuguese Foundation for Science and Technology) through national funds and when applicable co-financed financed by FEDER, under the Partnership Agreement PT2020 (Project CEMAPRE - UID/MULTI/00491/2019).}
 
 {\textbf{Special thanks} to the Superintendence of the INSS (Brazil) that provided the data.}
\bibliography{library}
\bibliographystyle{myapalike}

\vspace*{1cm}

Renata Gomes Alcoforado \\
ISEG \& CEMAPRE, Universidade de Lisboa,\\
Department of Accounting and Actuarial Sciences, Universidade Federal de Pernambuco\\
E-mail address: alcoforado.renata@ufpe.br

\vspace*{0.15cm}

Alfredo D. Egídio dos Reis \\
ISEG \& CEMAPRE, Universidade de Lisboa,\\
E-mail address: alfredo@iseg.ulisboa.pt

\vspace*{2cm}
\section*{Appendix A}

\begin{landscape}
\label{S:apend}
\appendix
	\begin{table}[]
		\begin{tabular}{cclccccccccc}
			\hline \hline
			& & & \multicolumn{9}{l}{Life Expectancy}\\
			\textbf{No.} & \textbf{UF} & \multicolumn{1}{c}{\textbf{Name of UF}} & \textbf{\begin{tabular}[c]{@{}c@{}}at Birth\\ Total\end{tabular}} & \textbf{\begin{tabular}[c]{@{}c@{}}at Birth\\ Male\end{tabular}} & \textbf{\begin{tabular}[c]{@{}c@{}}at Birth\\ Female\end{tabular}} & \textbf{\begin{tabular}[c]{@{}c@{}}After 60\\ Total\end{tabular}} & \textbf{\begin{tabular}[c]{@{}c@{}}After 60\\ Male\end{tabular}} & \textbf{\begin{tabular}[c]{@{}c@{}}After 60\\ Female\end{tabular}} & \textbf{\begin{tabular}[c]{@{}c@{}}After 65\\ Total\end{tabular}} & \textbf{\begin{tabular}[c]{@{}c@{}}After 65\\ Male\end{tabular}} & \textbf{\begin{tabular}[c]{@{}c@{}}After 65\\ Female\end{tabular}} \\ \hline
			\textbf{1}      & AL          & Alagoas             & 70.73                                                               & 66.88                                                              & 74.77                                                                & 80.6                                                                & 78.6                                                               & 82.4                                                                 & 82.2                                                                & 80.4                                                               & 83.7                                                                 \\
			\textbf{2}      & AM          & Amazonas            & 74.26                                                               & 71.23                                                              & 77.45                                                                & 80.5                                                                & 78.8                                                               & 82.1                                                                 & 81.9                                                                & 80.5                                                               & 83.4                                                                 \\
			\textbf{3}      & BA          & Bahia               & 74.61                                                               & 71.4                                                               & 77.97                                                                & 81.7                                                                & 79.5                                                               & 83.8                                                                 & 83.1                                                                & 81.1                                                               & 84.9                                                                 \\
			\textbf{4}      & CE          & Ceará               & 73.51                                                               & 69.47                                                              & 77.75                                                                & 81.5                                                                & 79.8                                                               & 82.9                                                                 & 82.8                                                                & 81.4                                                               & 84.1                                                                 \\
			\textbf{5}      & MS          & Mato Grosso do Sul  & 76.29                                                               & 73.11                                                              & 79.63                                                                & 82.2                                                                & 80.3                                                               & 84                                                                   & 83.5                                                                & 81.8                                                               & 85.1                                                                 \\
			\textbf{6}      & ES          & Espírito Santo      & 76.28                                                               & 72.82                                                              & 79.91                                                                & 84.1                                                                & 82                                                                 & 86.1                                                                 & 85.3                                                                & 83.3                                                               & 87                                                                   \\
			\textbf{7}      & GO          & Goiás               & 75.88                                                               & 72.67                                                              & 79.25                                                                & 81.3                                                                & 79.9                                                               & 82.6                                                                 & 82.6                                                                & 81.5                                                               & 83.6                                                                 \\
			\textbf{8}      & MA          & Maranhão            & 71.48                                                               & 67.69                                                              & 75.45                                                                & 80.4                                                                & 78.3                                                               & 82.4                                                                 & 82                                                                  & 80.2                                                               & 83.8                                                                 \\
			\textbf{9}      & MT          & Mato Grosso         & 75.66                                                               & 72.19                                                              & 79.3                                                                 & 81.4                                                                & 80                                                                 & 83                                                                   & 82.8                                                                & 81.6                                                               & 84.1                                                                 \\
			\textbf{10}     & MG          & Minas Gerais        & 76.96                                                               & 73.66                                                              & 80.42                                                                & 83.1                                                                & 81.6                                                               & 84.5                                                                 & 84.3                                                                & 83                                                                 & 85.5                                                                 \\
			\textbf{11}     & PA          & Pará                & 74.54                                                               & 71.58                                                              & 77.65                                                                & 80.55                                                               & 78.9                                                               & 82.2                                                                 & 82                                                                  & 80.5                                                               & 83.5                                                                 \\
			\textbf{12}     & PB          & Paraíba             & 72.54                                                               & 69.13                                                              & 76.12                                                                & 81.2                                                                & 79.7                                                               & 82.6                                                                 & 82.6                                                                & 81.2                                                               & 83.6                                                                 \\
			\textbf{13}     & PR          & Paraná              & 76.72                                                               & 73.55                                                              & 80.04                                                                & 82.7                                                                & 80.9                                                               & 84.3                                                                 & 83.8                                                                & 82.3                                                               & 85.2                                                                 \\
			\textbf{14}     & PE          & Pernambuco          & 71.97                                                               & 68.55                                                              & 75.56                                                                & 81.2                                                                & 79.3                                                               & 82.7                                                                 & 82.5                                                                & 80.9                                                               & 83.8                                                                 \\
			\textbf{15}     & PI          & Piauí               & 72.52                                                               & 69.44                                                              & 75.75                                                                & 79.8                                                                & 77.8                                                               & 81.6                                                                 & 81.3                                                                & 79.6                                                               & 82.8                                                                 \\
			\textbf{16}     & RJ          & Rio de Janeiro      & 75.88                                                               & 71.86                                                              & 80.09                                                                & 82.4                                                                & 80.2                                                               & 84.1                                                                 & 83.7                                                                & 81.7                                                               & 85.2                                                                 \\
			\textbf{17}     & RN          & Rio Grande do Norte & 73.65                                                               & 69.96                                                              & 77.52                                                                & 82.4                                                                & 80.3                                                               & 84.3                                                                 & 83.7                                                                & 81.7                                                               & 85.2                                                                 \\
			\textbf{18}     & RS          & Rio Grande do Sul   & 77.26                                                               & 73.78                                                              & 80.92                                                                & 83.1                                                                & 80.8                                                               & 85                                                                   & 84.2                                                                & 82.2                                                               & 85.9                                                                 \\
			\textbf{19}     & SC          & Santa Catarina      & 77.49                                                               & 74.33                                                              & 80.8                                                                 & 83.9                                                                & 81.6                                                               & 86                                                                   & 85                                                                  & 82.8                                                               & 86.8                                                                 \\
			\textbf{20}     & SP          & São Paulo           & 76.79                                                               & 72.94                                                              & 80.83                                                                & 83.2                                                                & 81.2                                                               & 84.9                                                                 & 84.4                                                                & 82.6                                                               & 85.9                                                                 \\
			\textbf{21}     & SE          & Sergipe             & 74.04                                                               & 70.73                                                              & 77.52                                                                & 80.6                                                                & 78.6                                                               & 82.4                                                                 & 82.1                                                                & 80.3                                                               & 83.6                                                                 \\
			\textbf{22}     & DF          & Distrito Federal    & 77.45                                                               & 73.94                                                              & 81.13                                                                & 83.3                                                                & 81.1                                                               & 85.2                                                                 & 84.3                                                                & 82.3                                                               & 85.9                                                                 \\
			\textbf{23}     & AC          & Acre                & 74.1                                                                & 71.39                                                              & 76.95                                                                & 81.6                                                                & 79.9                                                               & 83.4                                                                 & 83.1                                                                & 81.5                                                               & 84.8                                                                 \\
			\textbf{24}     & AP          & Amapá               & 73.28                                                               & 69.55                                                              & 77.19                                                                & 81.8                                                                & 80.5                                                               & 83.2                                                                 & 83.2                                                                & 81.9                                                               & 84.4                                                                 \\
			\textbf{25}     & RO          & Rondônia            & 73.96                                                               & 71.13                                                              & 76.94                                                                & 79.5                                                                & 78.3                                                               & 81                                                                   & 81                                                                  & 79.9                                                               & 82.2                                                                 \\
			\textbf{26}     & RR          & Roraima             & 72.94                                                               & 70.33                                                              & 75.68                                                                & 79.9                                                                & 79                                                                 & 80.8                                                                 & 81.3                                                                & 80.4                                                               & 82.3                                                                 \\
			\textbf{27}     & TO          & Tocantins           & 74.01                                                               & 71.58                                                              & 76.57                                                                & 81.3                                                                & 80.1                                                               & 82.7                                                                 & 82.7                                                                & 81.6                                                               & 83.8                                                               \\ 
			\hline \hline \\
		\end{tabular}
\caption{UF's with codes and life expectancies}
	\label{T:codigos}
	\end{table}
\end{landscape}

\begin{landscape}
	\begin{figure}[h]
		\center
		\includegraphics[width=24cm]{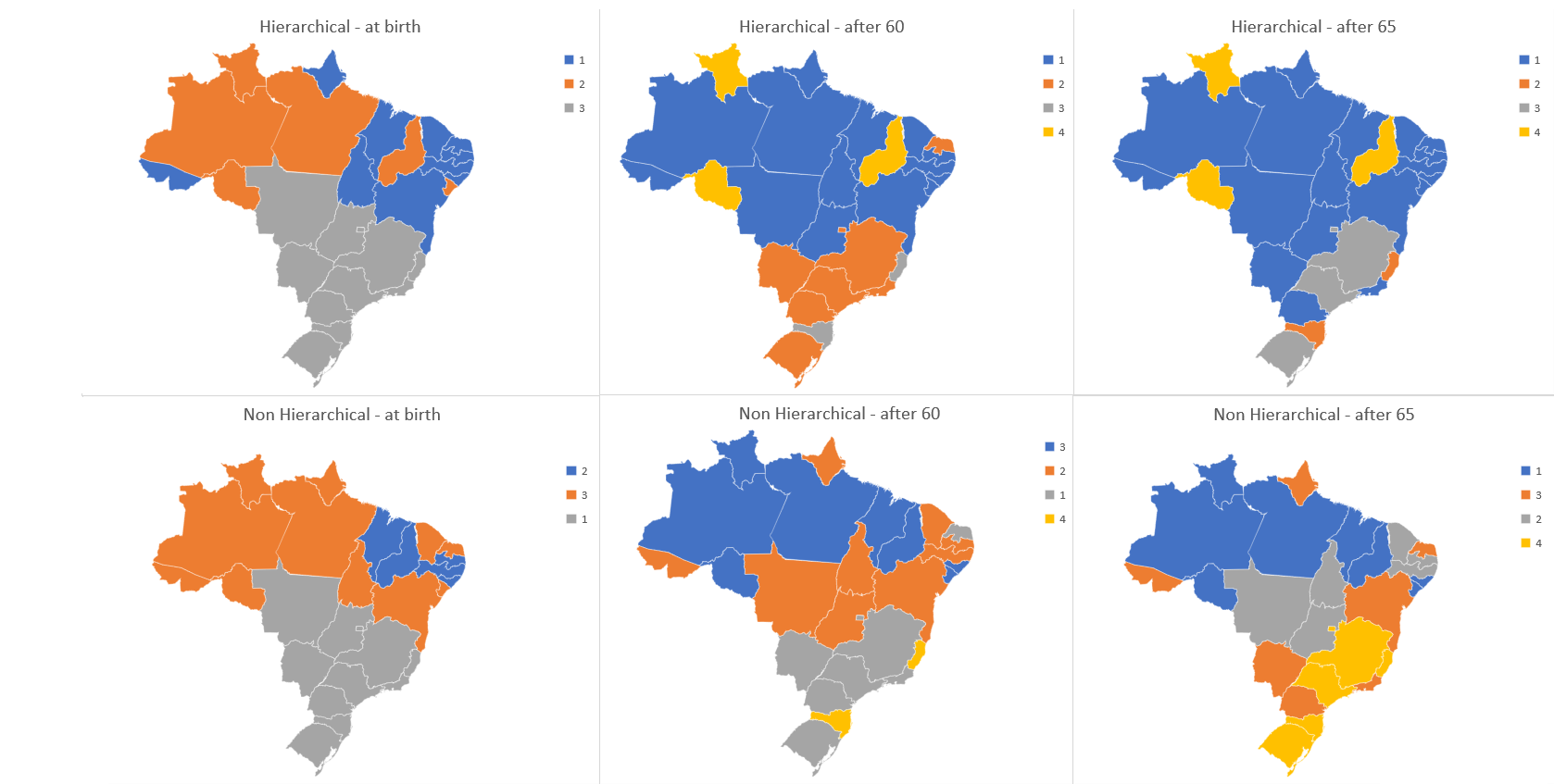}
		\caption{Spacial clusters}
		\label{F:clusters}
	\end{figure}
\end{landscape}

\begin{figure}[h]
	\center
	\includegraphics[width=11cm]{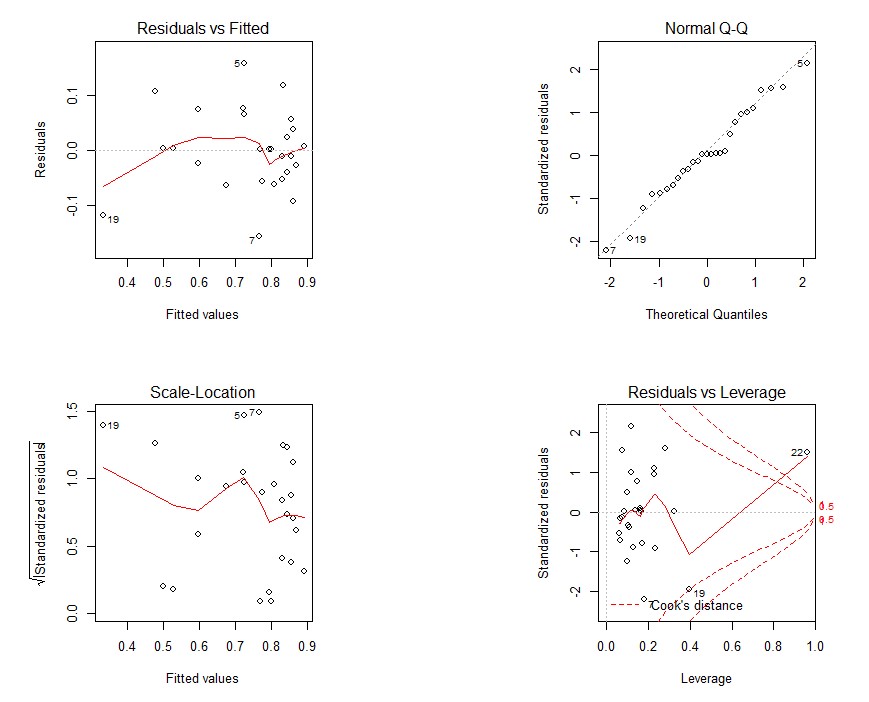}
	\caption{Fit for the Social Support for the Elderly and Disabled}
	\label{F:mod}
\end{figure}

\begin{figure}[h]
	\center
	\includegraphics[width=11cm]{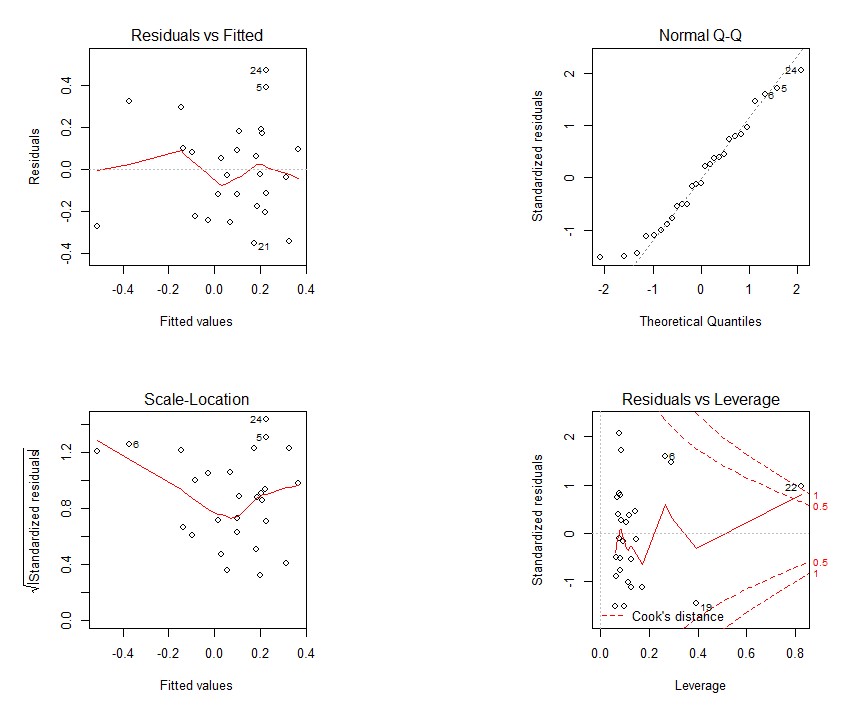}
	\caption{Fit for the Social Support for the Elderly}
	\label{F:mod1}
\end{figure}

\begin{figure}[]
	\center
	\includegraphics[width=11cm]{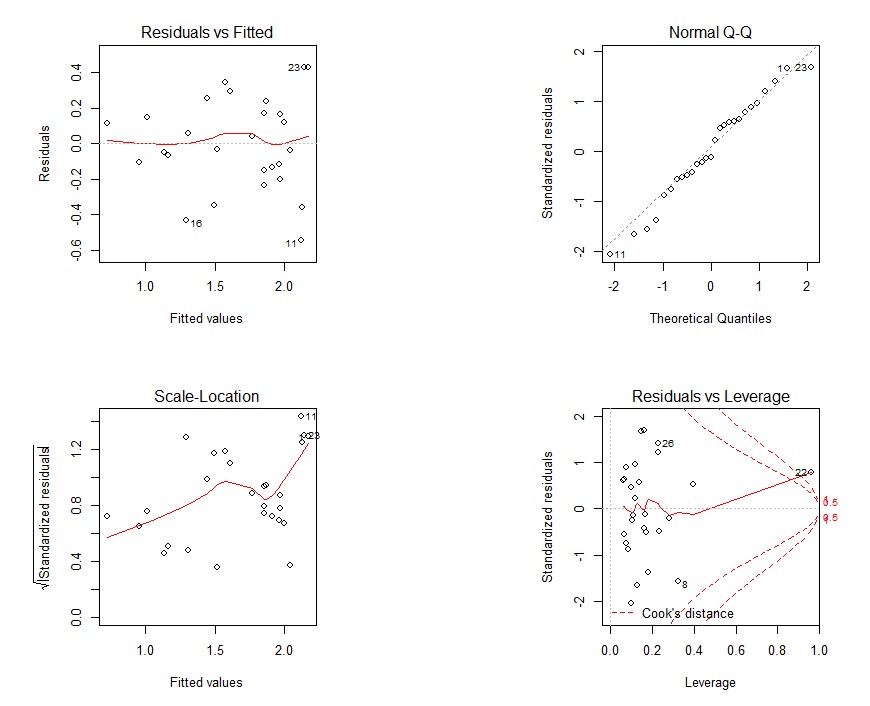}
	\caption{Fit for the Social Support for the Disabled}
	\label{F:mod2}
\end{figure}
\end{document}